\pdfoutput=1 

\documentclass[sigconf,10pt]{acmart}
\usepackage{hyperref}
\usepackage{tikz}
\usepackage{amsmath,bm}
\usepackage{filecontents}
\usepackage{graphicx}
\usepackage{breakurl}
\usepackage{breqn,xspace}
\usepackage{subfigure, bm}
\usepackage{multirow}
\usepackage{threeparttable}
\usepackage{mathrsfs}
\usepackage{pifont}
\usepackage{caption}
\usepackage{subfigure}
\usepackage[noend]{algorithm2e}
\usepackage{dblfloatfix}
\usepackage{lipsum,multicol}

\newcommand{\cm}[1]{{\color{red}\bf[#1]}} 

\newcommand{\system}{\textsc{Medusa}\xspace}

\setcopyright{acmcopyright}

\copyrightyear{2025}
\acmYear{2025}
\setcctype{by}
\acmConference[ACM MOBICOM '25]{The 31st Annual International Conference
on Mobile Computing and Networking}{November 4--8, 2025}{Hong Kong, China}
\acmBooktitle{The 31st Annual International Conference on Mobile Computing
and Networking (ACM MOBICOM '25), November 4--8, 2025, Hong Kong,
China}\acmDOI{10.1145/3680207.3723461}
\acmISBN{979-8-4007-1129-9/2025/11}

\begin{document}


\title[Scalable Multi-View Biometric Sensing in the Wild with Distributed MIMO Radars]{\system: Scalable Multi-View Biometric Sensing in the Wild with Distributed MIMO Radars} 

\author{
Yilong Li\textsuperscript{1}, 
Ramanujan K Sheshadri\textsuperscript{2},
Karthik Sundaresan\textsuperscript{3},
Eugene Chai\textsuperscript{2},
Yijing Zeng\textsuperscript{1},
Jayaram Raghuram\textsuperscript{1},
Suman Banerjee\textsuperscript{1}
}

\affiliation{
\textsuperscript{1} University of Wisconsin-Madison \quad
\textsuperscript{2} Nokia Bell Labs \quad
\textsuperscript{3} Georgia Institute of Technology \quad
\country{\{yilong,yijingzeng,suman\}@cs.wisc.edu \quad karthik@ece.gatech.edu \quad
jraghuram@wisc.edu \quad \{ram.sheshadri,eugene.chai\}@nokia-bell-labs.com \quad}
}



\renewcommand{\shortauthors}{Y.~Li et al.}
    
\begin{abstract}
Radio frequency (RF) techniques have shown promise for continuous contactless healthcare applications. However, real-world indoor environments pose challenges for existing systems, which may struggle to detect subtle physiological signals.
%
%
%
This paper proposes \system, a novel wireless vital-sign sensing system designed for multi-view setups. It enables users to deploy distributed Multiple Input Multiple Output (MIMO) arrays into their daily living environments, facilitating vital-sign sensing in real-world settings.
Unlike most existing single Commercial Off-The-Shelf (COTS) radar-based systems that operate
under controlled settings
\system's primary novelty lies in the design of a first-of-its-kind flexible \textit{multi-view} vital sign sensing system that is view-agnostic, pose-agnostic, contactless, and can sense basic human vitals with good accuracy. 
%
Through our well-engineered hardware and software co-design, \system enables real-time processing of large distributed MIMO arrays, while balancing the tradeoff between  Signal-to-Noise Ratio (SNR) and spatial diversity gain across each of its four distributed $4 \times 4$ sub-arrays for increased robustness. This is achieved using our \textit{novel unsupervised learning} model 
which effectively recovers vital sign waveforms by decomposing the received signals. 
%
Extensive evaluations with 21 participants demonstrate \system’s spatial diversity gain for real-world vital-sign monitoring, enabling free movement and orientation of subjects in both familiar and unfamiliar indoor environments.
\end{abstract}

\maketitle

\vspace{-1.5ex}
\section{Introduction}
\label{intro}
\vspace{-0.5ex}
Over the last ten years, digital wellness has become increasingly popular, especially in the field of passive health monitoring (PHM)
without the need for on-body devices. This trend supports a range of applications, such as remote physical rehabilitation, vital sign monitoring, and fall detection in indoor settings~\cite{adib:2015,mobi2sense,movifi,morefi,gong_UbiComp2021}.
%
These systems offer numerous advantages to both users and physicians. Users are unburdened from the correct usage and maintenance of sensor devices, allowing them to participate freely in daily activities. On the other hand, physicians can implement continuous patient-monitoring protocols and facilitate proactive management of their health.

There have been a lot of recent advancements in wireless-driven PHM 
, ranging from simple respiration rate monitoring (e.g., ~\cite{adib:2015}) to more advanced biometric sensing solutions using various technologies (e.g., UWB~\cite{morefi,movifi,mobi2sense,gong_UbiComp2021}, mmWave~\cite{ha2020contactless,gong_UbiComp2021}, FMCW Radar~\cite{adib:2015}, Wi-Fi~\cite{fullbreathe,NiWiFi2022,liu2015}). While prevalent contact sensing technologies, including wristbands and chest vests, allow vital sign monitoring, such devices present limitations in prolonged use due to discomfort arising from contact with the skin, and lack of robustness in detecting moving human subjects. 

\begin{figure}[ht!]
\vspace{-0.5ex}
    \centering
  \includegraphics[width=0.95\columnwidth]{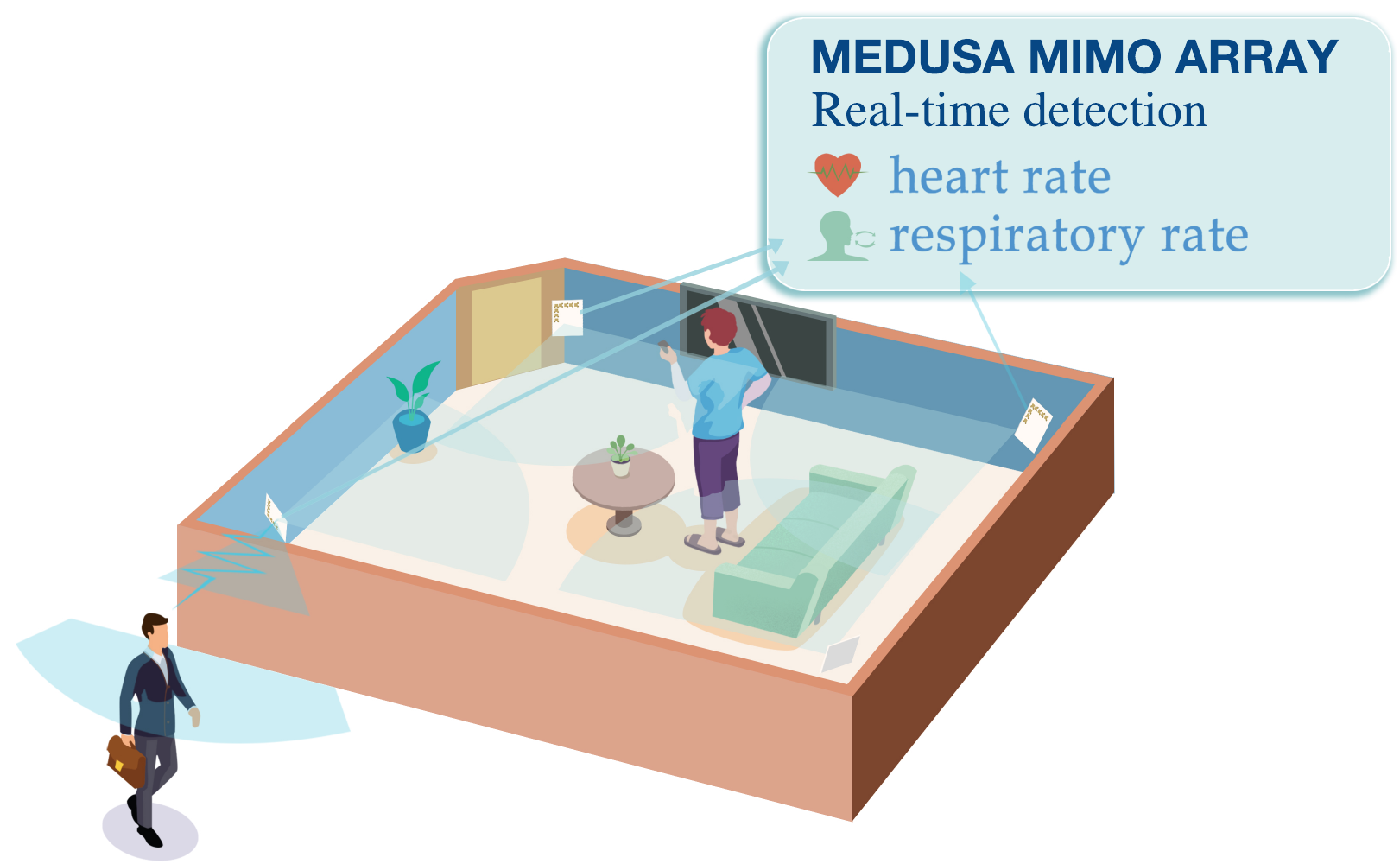}
  \vspace{-1ex}
    \caption{Vital-sign monitoring of human subjects in their daily habitats enables free movement and multiple subjects.}
        \label{fig:intro}
        \vspace{-2.2ex}
\end{figure}

Despite their potential, existing radar-based vital-sign monitoring systems face challenges and decreased performance in real-world environments. \textbf{(i)} Lack of Robustness to Blockage and Motion: furniture and other environmental obstacles can block radar signals in real-world settings. Also, body parts may obstruct the radar path during subject movement. 
Commercial off-the-shelf (COTS) mmWave radar devices face limitations in penetration, which hinders their ability to reliably distinguish the desired signals in the real world. \textbf{(ii)} Limited Applicability: Existing radar-based vital-sign sensing systems operate under controlled conditions due to limited antenna arrays, requiring subjects to face the radar directly. They allow only limited motion, either stationary or small movements, within a narrow field-of-view (FoV). \textbf{(iii)} Limited Coverage from Single-View Sensing: single-view vital-sign sensing systems, whether using a single Tx-Rx COTS radar or a single-perspective MIMO configuration, can only capture partial information from a human subject, especially in NLoS. Single-view systems are particularly vulnerable to occlusions caused by the subject's body parts when the subject is not directly facing the radar. These occlusions can obstruct the radar signals, resulting in inaccurate detection.

The primary question posed in this research is: {\em Is it feasible to transition radar sensing of biometric signals from restrictive deployments to practical \textbf{\textit{in-the-wild}} operation, allowing for the monitoring of subjects in their daily life environment without sacrificing the sensing performance?} 

We propose a novel RF-based vital sign monitoring system using distributed MIMO arrays, where the \textbf{multi-view} setup isolates vital signs from noise and motion interference, enabling detection in ambulatory settings with unrestricted movement and orientation.
The core intuition behind \system is to leverage the diverse signal paths and perspectives enabled by spatially distributed MIMO radar arrays for ``multi-view sensing'', as shown in Fig.~\ref{fig:intro}. 
%
%
Unlike existing ``single-view'' systems that can only capture partial vital-sign information as people change orientations, our distributed MIMO approach allows the gathering of multi-aspect data.

While distributed MIMO arrays contribute to a throughput improvement in communication systems, it can potentially deliver a more profound impact for wireless sensing -- going from constrained operation to enabling practical ``in-the-wild'' sensing operation that is robust and immune to varying target locations, orientations, blockages, and mobility. 
%
Towards realizing this objective, we present \system -- a large systems effort that addresses the following key challenges, leading to its contributions:\footnote{The open-source design files and repository are available on website: https://jimmy-yilong-li.github.io/.}

\begin{figure*}[ht!]
\vspace{-0.5ex}
    \setlength\abovecaptionskip{3pt}
     \centering
     \subfigure[System Platform Architecture]{
         \includegraphics[width=1\columnwidth]{figures/intro_arch.pdf}
         \label{fig:intro_arch}
     }
      \subfigure[Unsupervised Learning]{
         \includegraphics[width=0.85\columnwidth]{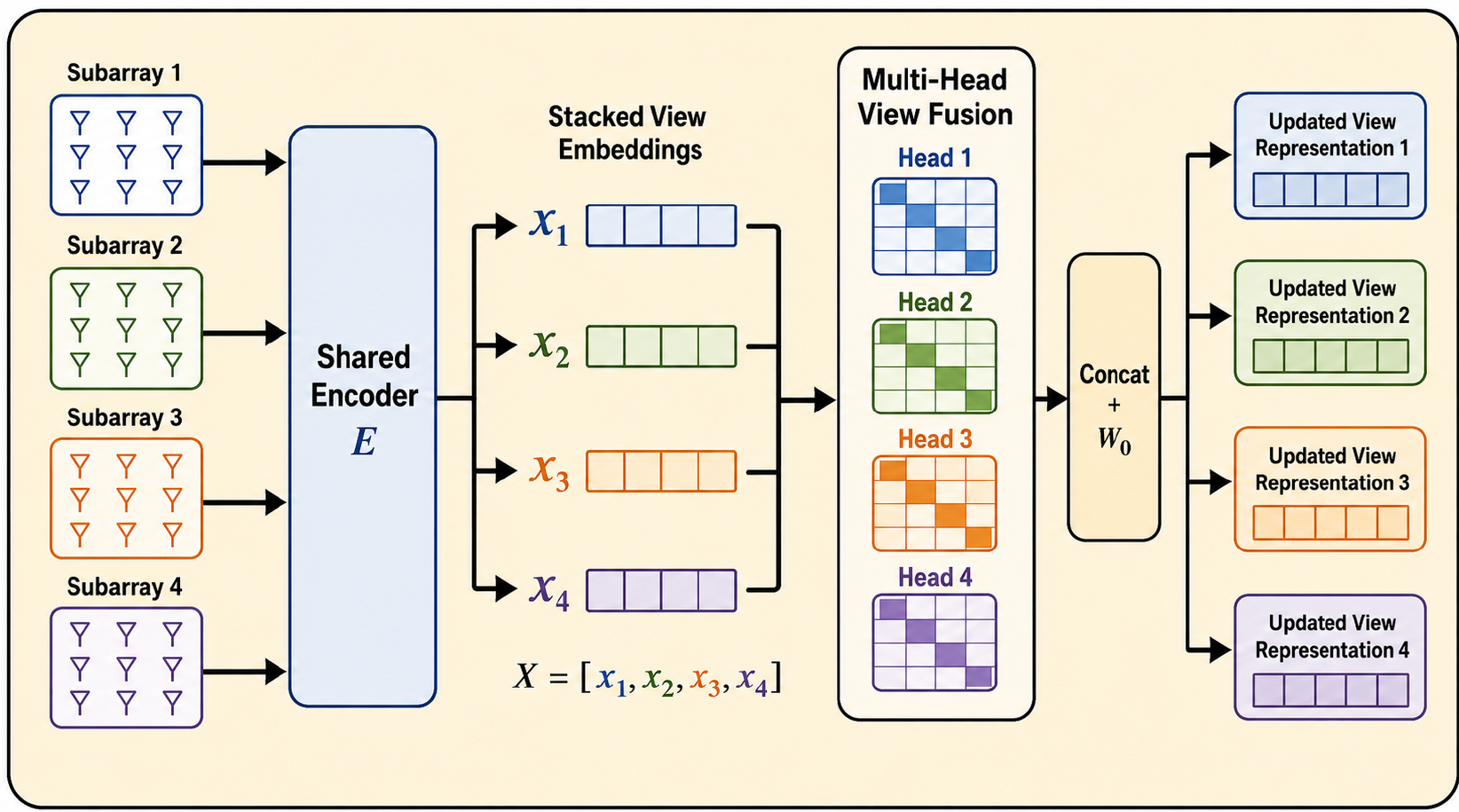}
         \label{fig:intro_model}
     }
     \caption{System architecture diagram of \system.}
        \label{fig:introarch}
        \vspace{-2ex}
\end{figure*}

\noindent \textbf{1. Distributed Radar System:} How do we create a practical, distributed MIMO radar system that has sufficient gain to operate in real-time and reliably for practical room-scale deployments? While UWB and mmWave radar systems offer high resolution (1-4 GHz bandwidth), radar sensing offers a very limited operational range (less than 3m) with today's systems that have limited arrays (e.g., $4 \times 3$ elements virtual mmWave array~\cite{TImmWave1443}). 
This challenge is exacerbated by interference from multiple radars transmitting simultaneously without tight synchronization, a difficult feat in conventional distributed radar systems.

\textbf{Contribution 1.} \system builds a \textit{first-of-its-kind} 16 UWB radar elements MIMO system that enables flexible, coherent, distributed vital sign sensing in a multi-view way (shown in Fig.~\ref{fig:introarch}). The wireless synchronization facilitates easy deployment and ensures robustness across diverse target and environment configurations. mmWave radar suffers from signal blockage in NLoS scenarios. In contrast, UWB's better penetration capability (3-10 GHz) compared to mmWave, coupled with the large antenna gain in \system, allows us to detect targets as far as 6m, even in NLoS reliably.

\noindent \textbf{2. Balancing MIMO SNR Gain vs. Spatial Diversity Gain:}  
The theory of distributed MIMO radar is well understood~\cite{biglieri2007mimo}. 
However, dispersing the MIMO array into various sub-arrays can amplify spatial diversity gain, even if the SNR of individual sub-arrays is reduced. 
%
We revisit the trade-off between antenna gain (SNR) and diversity gain (outage probability) in sensing. A single 256-element virtual MIMO array $16 Tx \times 16 Rx$) maximizes gain from one viewpoint but is vulnerable to subjects' orientation, location, NLoS, blockages, and motion variations. Conversely, distributing sub-arrays across multiple locations enhances robustness to ``in-the-wild'' configurations with reduced antenna gain and coverage per sub-array. The challenge is efficiently managing this tradeoff to ensure accurate and robust sensing across varied target and environmental conditions.

\textbf{Contribution 2.} Through extensive experimental and motivational characterization,
\system identifies an ideal operating point, wherein one sub-array per quadrant around the target area strikes the most effective balance -- any further distribution only reduces the gain per sub-array without adding appreciably to the diversity gain. In contrast, less diversity diminishes the  
robustness of sensing performance. In particular, \system employs four $4\times4$ sub-arrays, each with 16 virtual MIMO elements, to establish its distributed radar system for rooms. This configuration also offers flexibility to adapt to other room shapes as needed.


\noindent \textbf{3. Multi-view Information Fusion and Signal Extraction:}  While \system enables optimal balanced SNR sensing signals and multi-view information in dynamic environments, it is still a hard challenge to isolate the vital sign signals mixed with numerous other reflections (distortions from human motion, multipath, etc.) in a non-linear manner. Additionally, extracting desired signals from different angles of the subject requires synchronizing and fusing the collected vital sign signals.
Machine learning offers a scalable approach to extract our desired signal, given its recent success in several wireless sensing problems~\cite{adib:2015,movifi,morefi,gong_UbiComp2021,dinadeepbreath}. However, these learning models are limited to controlled settings and single-view COTS radar data, lacking the ability to process multi-view information. 
Furthermore, every environment leaves its artifacts in a dataset, significantly affecting the model's generality when tested on unseen environments with different configurations.


\textbf{Contribution 3.} \system employs an unsupervised approach to extract target breathing signals across various environments, accommodating natural movement during everyday tasks. To harness the diversity of distributed sub-arrays, it integrates a multi-head attention layer that dynamically attends to different sub-arrays based on the target’s estimated location, orientation, and environmental blockages. While Independent Component Analysis (ICA) and multi-head attention are well-established individually, our work uniquely combines these techniques for multi-view, non-contact physiological sensing. This approach enhances spatial diversity in multi-view antenna arrays, improving signal separation and detection accuracy.

Our comprehensive evaluation with 27 participant subjects reveals that \system improves median respiration measurement by over 20\% accuracy compared to prior art and baselines in practical indoor environments,  characterized by varying subject locations and orientations, and sustains errors under 5\% even in NLoS (obstacles) conditions, where other approaches falter. Further, leveraging the diversity benefits of its distributed platform allows its model to (generalize) sustain performance accuracy even in unseen environments, subject mobility, and multiple subjects.

To the best of our knowledge, \system is the first-its-kind real-time vital sign monitoring system designed to function effectively in dynamic real-world environments,
which is pioneering in incorporating multi-view information fusion into wireless vital sign sensing.

To date, \system has been evaluated only on healthy subjects (per our IRB protocol). As the system evolves, we plan to transition to clinical trials with patients and explore diagnostic applications using vital sign data. In Section~\ref{sec:discuss}, we discuss that \system’s respiration rate accuracy falls within clinically acceptable error ranges, and heart rate detection is largely within threshold in short range, demonstrating significant potential for robust monitoring.

\vspace{-1.5ex}
\section{Motivation}
\label{sec:motivate}
\vspace{-0.5ex}

\begin{figure}[htb]
\vspace{-0.5ex}
  \centering
  \includegraphics[width=1\columnwidth]{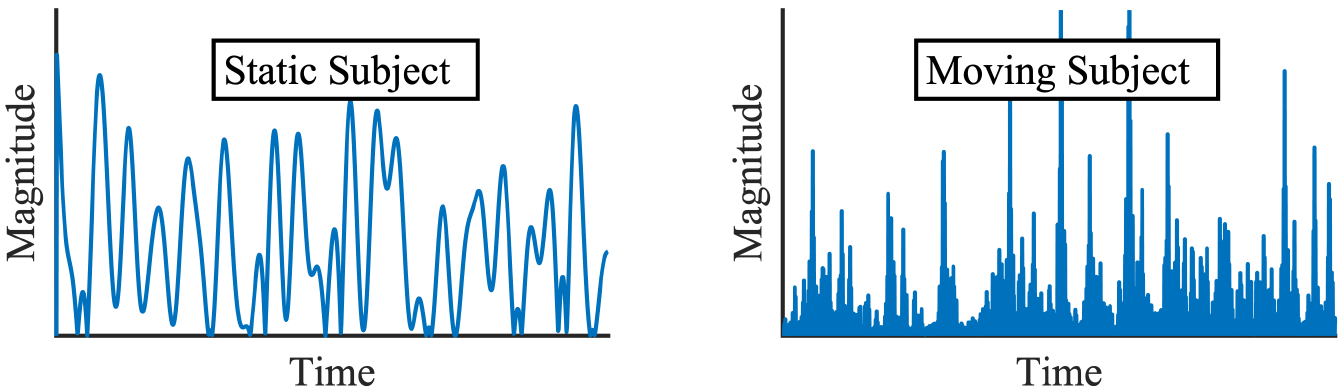}
  \vspace{-1ex}
  \caption{Respiration signals by vibration of COTS radar: Static subject vs. Moving subject.}
  \label{fig:waveformstaticandmoving}
  \vspace{-2ex}
\end{figure}
In this section, we 
empirically study the impact of human movement and orientation on wireless vital sign monitoring and motivate the advantages of \emph{distributed} MIMO array-based sensing~(multi-view) over traditional co-located MIMO antenna systems~(single-view).


\vspace{-1.6ex}
\subsection{Case for Distributed MIMO Sensing}
\begin{figure}[!htb]
\vspace{-0.5ex}
\centering
\includegraphics[width=0.9\columnwidth]{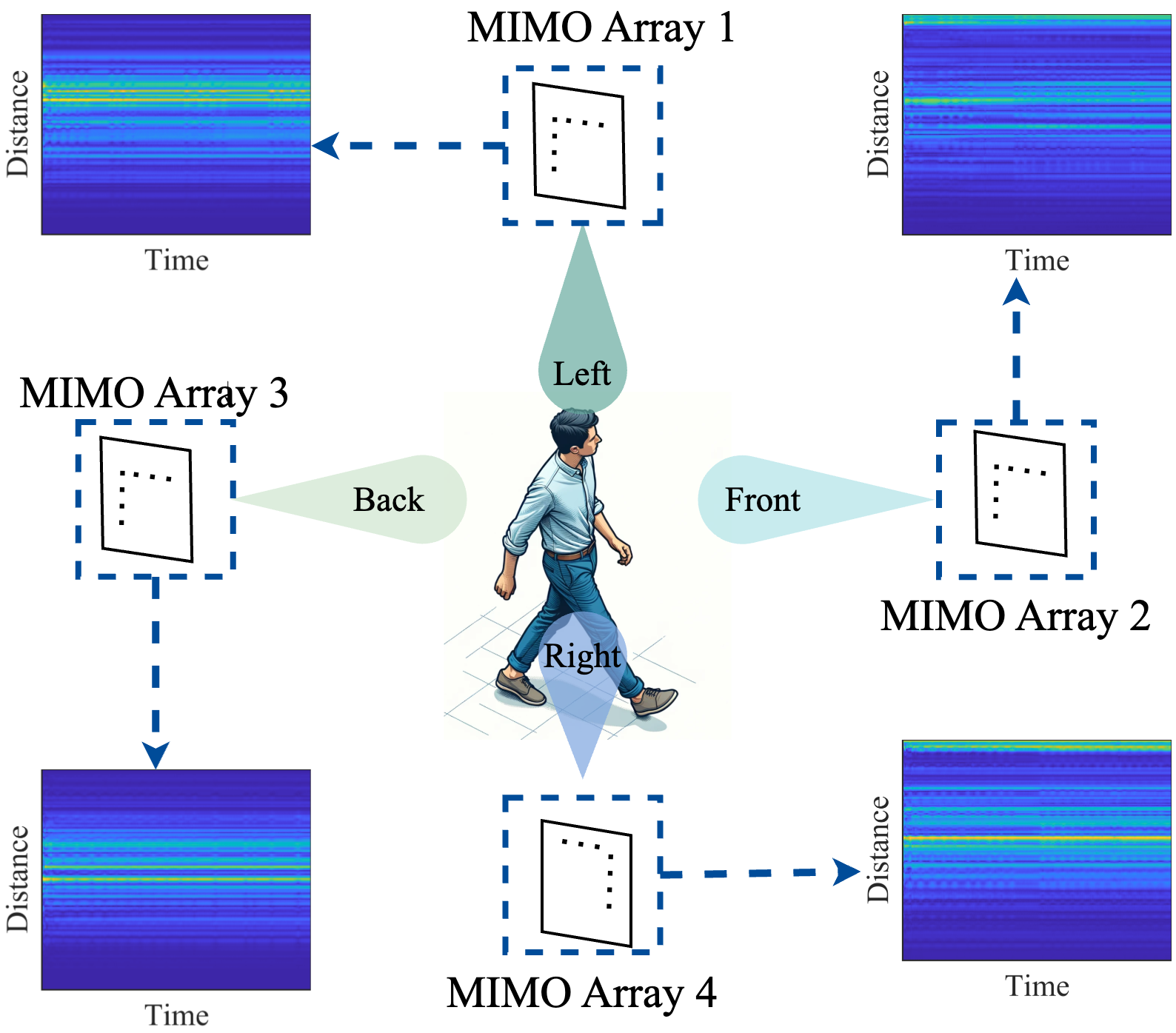}
\vspace{-2ex}
\caption{Spatial diversity from distributed MIMO arrays~(1, 2, 3, 4) in multi-view.} 
\label{fig:multiviewframe}
\vspace{-2.5ex} 
\end{figure}

\vspace{-0.5ex}
%
The high-level idea of \system is to strike a tradeoff between MIMO antenna gain for SNR vs. the spatial diversity gain~(multi-view) for a dynamic environment. As depicted in Fig. \ref{fig:multiviewframe}, the radar data from multiple angles provides diverse ``viewpoints'' of the target. In real environments, furniture and obstacles may obstruct some signal paths. However, as subjects move, it's improbable that all directions are simultaneously blocked.
Multi-view information from distributed MIMO radar arrays can prevent performance drops in traditional single-view sensing systems. Sensing information from angles beyond the front view of the human subject provides complementary data, thereby achieving diversity gain through these different "viewpoints."  Our motivational experiments illustrate how spatial diversity solves the challenges discussed before.

%

\begin{figure}[htb!]
\vspace{-2ex}
     \centering
      \subfigure[LOS]{
        \centering
         \includegraphics[width=0.45\columnwidth]{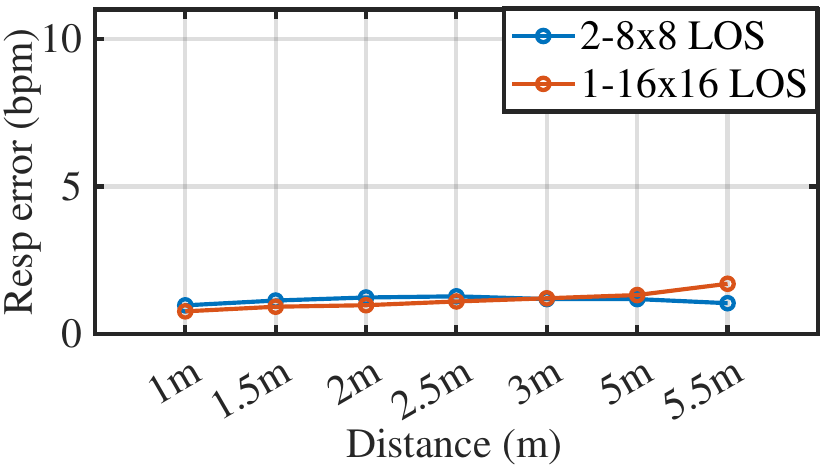}
         \label{fig:bpmcomparisonNLOSa}
     }
     \subfigure[NLoS]{
        \centering
         \includegraphics[width=0.45\columnwidth]{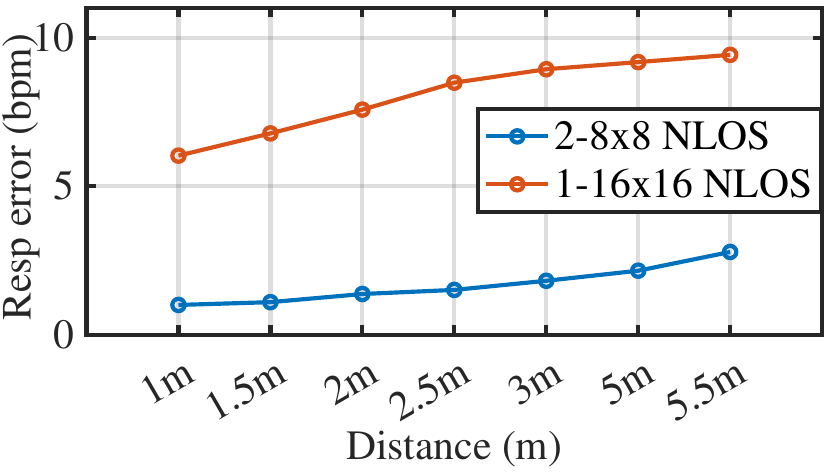}
         \label{fig:bpmcomparisonNLOSb}
     }
     \vspace{-1.5ex}
     \caption{Single-view $16\times16$ MIMO vs. two $8\times8$ MIMO radars in LOS and NLOS scenarios with subjects facing the radar.}
        \label{fig:bpmcomparisonNLOS}
        \vspace{-2.2ex}
\end{figure}

\noindent \textbf{Lack of Robustness to Blockage and Movement:}
\label{sec:nlos_impact}
Environmental structures and body movements can obstruct radar signals, distorting respiration patterns for both static and moving subjects (see Fig.~\ref{fig:waveformstaticandmoving}) in real-world settings. Figure~\ref{fig:bpmcomparisonNLOS} shows that splitting our \(16\times16\) array into two \(8\times8\) arrays placed at 90 degrees improves performance in obstructed environments. In contrast, a single \(16\times16\) array, when blocked, suffers from significant degradation with error rates exceeding 25\%.

\begin{figure}[ht!]
\vspace{-0.5ex}
    \setlength\abovecaptionskip{3pt}
     \centering
     \subfigure[SNR vs. Distance (Azimuth $0^\circ$)]{
         \includegraphics[width=0.46\columnwidth]{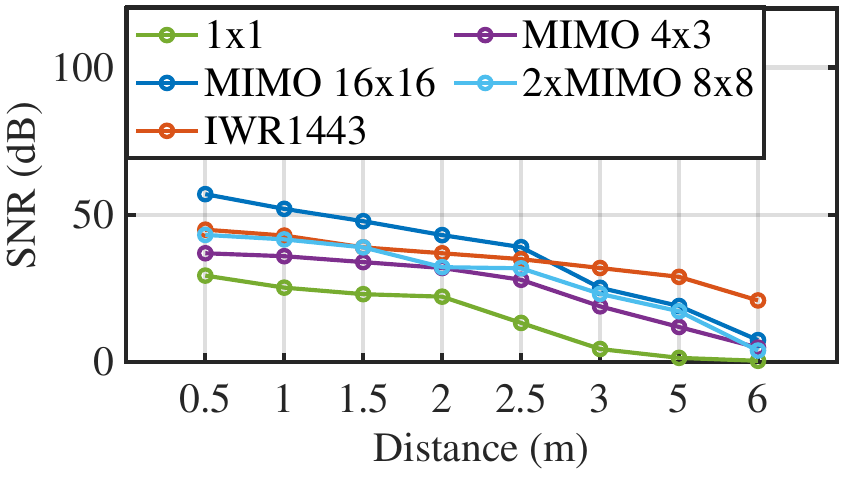}
         \label{fig:multiplevssingleSNRDistance}
     }
      \subfigure[SNR vs. FoV (Distance 1m)]{
         \includegraphics[width=0.46\columnwidth]{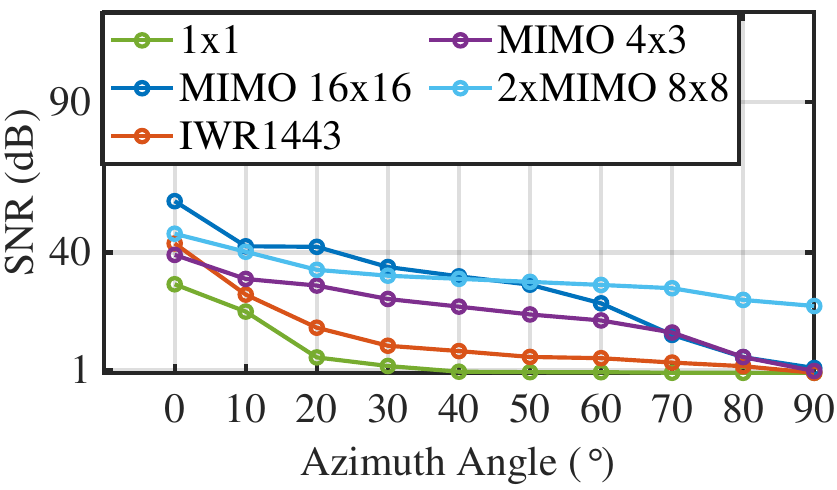}
         \label{fig:multiplevssingleSNRAngle}
     }
     \caption{Distance and FoV measurement ($16\times16$ MIMO, $4\times3$ MIMO, $1 \times 1$ COTS UWB radar, TI IWR1443 mmWave radar and two $8 \times 8$ MIMO).}
        \label{fig:multiplevssingleSNR}
        \vspace{-2ex}
\end{figure}

\noindent \textbf{Limited Applicability:}
\label{sec:limted}
COTS UWB radar sensors used in previous studies~\cite{movifi,morefi,mobi2sense} have a SISO configuration, limiting the operating range. MIMO radars offer superior spatial resolution and higher SNR gain compared to SISO systems, offering increased operating range across both distance and angular dimensions. This is evident in  
the $16 \times 16$ (256-element virtual) radar, as shown in Fig. \ref{fig:multiplevssingleSNR}.
It achieves a substantial 10-30 dB SNR gain, particularly when the target directly faces the antennas~(Azimuth $0^\circ$), enabling practical operational ranges exceeding 6m. However, the performance of co-located MIMO systems deteriorates when the target is outside the field-of-view (FoV) in Fig~\ref{fig:multiplevssingleSNRAngle}, resulting in a degraded SNR.
While the mmWave MIMO radar delivers a good SNR when the target directly faces it (Azimuth $0^\circ$), its performance degrades significantly out of FoV.

\begin{figure}[ht!]
\vspace{-0.5ex}
    \setlength\abovecaptionskip{3pt}
     \centering
     \subfigure[One single-view $16\times16$ array vs. two $8\times8$ arrays: Respiration errors while facing the antennas.]{
         \includegraphics[width=0.46\columnwidth]{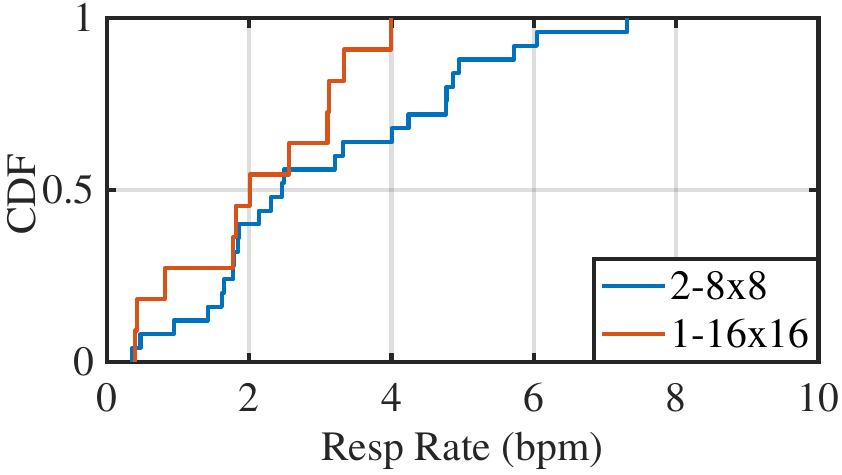}
         \label{fig:MIMOfront16vs8}
     }
      \subfigure[One single-view $16\times16$ array vs. two $8\times8$ arrays: Respiration errors while Facing other orientations.]{
         \includegraphics[width=0.46\columnwidth]{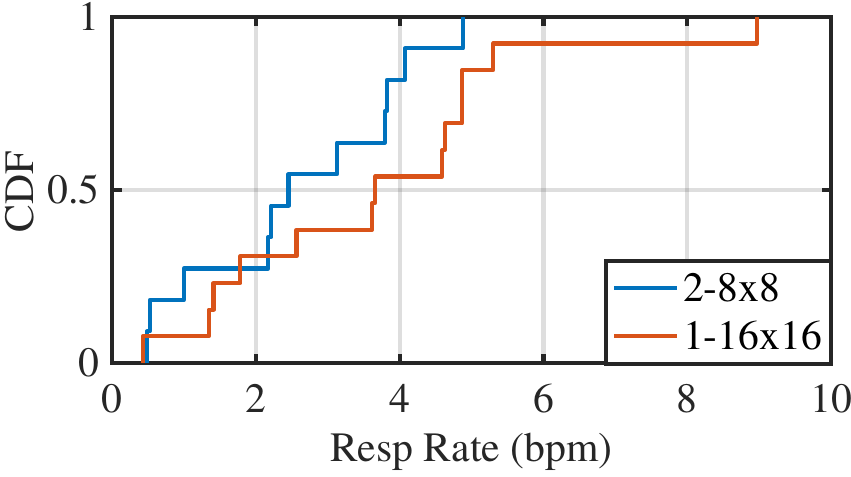}
         \label{fig:MIMOorients16vs8}
     }
     \caption{Single-view (one $16\times16$ radar) vs. multiple-view radar (two $8 \times 8$).}
        \label{fig:MIMO16vs8}
        \vspace{-3ex}
\end{figure}

\noindent \textbf{Limited Coverage of Single-View Radars:} 
\label{sec:distributed-mimo}
%
While conventional single-view (co-located) MIMO arrays offer high SNR and coverage, their ability to capture target reflections diminishes when the target is not front-facing. In contrast, spatially distributed MIMO systems provide the necessary diversity for vital sign monitoring regardless of the target's orientation or distance. As shown in Fig.~\ref{fig:MIMO16vs8}, two \(8\times8\) arrays placed at $90^\circ$ achieve a median error under 5\% for a static, front-facing target, albeit with slightly reduced accuracy compared to a single \(16 \times 16\) array. Moreover, while blockage is unavoidable in co-located arrays (e.g., a refrigerator causing 6-7 dB attenuation), distributed arrays with multiple viewpoints increase the likelihood of maintaining an unobstructed signal path.

\vspace{-1.6ex}
\subsection{\system Design Challenges}
\vspace{-1ex}
While our experimental study motivates the need for a distributed MIMO radar for robust HVM across target and environment dynamics, realizing such a system is challenging. 
\textbf{\underline{\emph{Synchronization and Phase Coherency:}}} 
Ensuring synchronization of each radar element, each with its own TX/RX chain, is critical in active MIMO radar systems. This presents a significant hardware challenge, as the system must maintain adequate gain across all distributed subarrays while preventing interference and SNR degradation caused by unsynchronized reflections. Despite these difficulties, high-precision, over-the-air clock synchronization is essential for the scalable deployment and reliable operation of distributed MIMO arrays.
    

\textbf{\underline{\emph{Balancing SNR vs. Diversity:}}} 
Large MIMO arrays (e.g., \(16\times16\)) boost SNR and coverage but lack robustness in practical scenarios. Conversely, spatially distributing a large array into several smaller subarrays enhances diversity; however, excessive distribution reduces gain and coverage per aperture without significant diversity improvements. An optimal balance between coverage and sensing robustness is, therefore, essential for efficient MIMO subarray deployment.

    
\textbf{\underline{\emph{Vital Sign Signal Extraction:}}}  To realize the true potential of distributed MIMO sensing for robustness towards target and environment dynamics, we need a scalable approach that can leverage and effectively fuse signals from the distributed sub-arrays to accurately extract the human vital signal from a complex non-linear mixture with other undesired signals (e.g., human motion, multipath reflections) in real-time. 

\vspace{-1.5ex}
\section{Related Work}
\label{sec:related}



Real-world applications require robust vital sign monitoring systems that can adapt to variations in target position, range, orientation, and motion, as well as environmental factors such as line-of-sight (LoS) and non-line-of-sight (NLoS) blockages, ensuring reliable in-the-wild performance.

\textbf{\emph{RF-based human sensing:}} The research community has long investigated the use of RF signals for human sensing and the impact of human movement~\cite{Rethink_doppler_JSAC2022, Acousticcardiogram, Widar_Mobisys2018, MilliSonic, chiang20203d, SMARS, WiDetect,WiSpeed, VeCare}. Research has made significant advances in using RF to passively (contact-free) sense people and their vital signs (i.e., human breathing and heart rates). 


\underline{\emph{Radar sensing:}}
Researchers in ~\cite{ha2020contactless,Widar_Mobisys2018, MilliSonic, movifi,morefi,uwbinfocom2022,mobi2sense,octopus,gong_UbiComp2021} have leveraged the large channel bandwidths available in 
UWB and mmWave bands to passively capture human breathing and heart rate. However, with most of these solutions employing
either the COTS UWB radar~\cite{novelda} with a single Tx/Rx antenna that the operational range is limited to only 3-5m distance, and some of the work the mmWave radar array (4 Tx, 3 RX~\cite{TImmWave1443}) operating at a high (75GHz-77GHz) frequency, even in LoS 
and within a narrow FoV of $\approx$ 50deg, making these solutions
inadequate for practical room-scale deployments, especially with NLoS. Moreover, having a single radar additionally constrains the target to
be oriented in a particular direction (e.g., facing the radar) to measure the physiological movements accurately. 

On the other hand, solutions~\cite{li:2008,ferreras:2016,wang:2011} that use two single-antenna radars, or
a single radar with two antennas~\cite{tang:2017}, require the two elements to be placed at specific locations with respect to a human body (e.g., front and back of a target) to eliminate signal-blockages caused by human action and body orientation, providing the simplest form of diversity, but are unable to offer practical ranges or NLoS operation with single antenna elements. 
Other solutions~\cite{ossi:2014,adib:2015,phuc:2016,yang:2017,V2iFi,gong_UbiComp2021, deepbreath} that use COTS devices, while capable of measuring vitals of multiple targets within an enclosed space, can do so only when the targets are reasonably stationary and facing the radar in LoS. In \cite{gong_UbiComp2021}, the approach of using COTS mmWave sensor takes advantage of a well-established correlation between HR/RR and motion intensity, moving away from the traditional time-frequency analysis, which is often disrupted by motion. However, it faces challenges in NLoS conditions and multiple targets. 
Additionally, vital sign monitoring during sleep, as explored in studies like \cite{deepbreath, SMARS}, focuses on detecting physiological signals from stationary subjects. \cite{deepbreath} proposed a method utilizing linear ICA to decompose respiratory signals from multiple individuals who remain motionless, such as those sleeping in fixed positions.

\underline{\emph{Wi-Fi sensing:}} WiFi-based sensing solutions~\cite{fullbreathe,liu2015,farsense,NiWiFi2022, Rethink_doppler_JSAC2022, Widar_Mobisys2018} rely on Channel State Information (CSI) to capture the minute displacements on a human body.
However, due to CSI’s high sensitivity to environmental changes and the limited bandwidth of Wi-Fi sensing, these solutions require extensive calibration and fingerprinting, making them impractical for real-world deployments.

\textbf{\emph{Non-RF based health sensing:}} In addition to RF-based sensing, researchers have explored acoustic~\cite{wang:2019,wang:2018,xu:2019,spirosonic} and 
camera-based\cite{chen:2018,pai:2021,yu:2019} solutions to monitor breathing and heart rates, respectively passively. However, 
they require targets to be relatively static in LoS, facing the radar and are impacted by background noise and human motion.

To the best of our knowledge, existing solutions do not effectively account for both target movement and environmental dynamics simultaneously

\vspace{-2.1ex}
\section{System Design}
\label{sec:design}
\vspace{-0.5ex}
Medusa employs a modular design allowing flexible deployment and optimal performance across diverse environments. Multiple radar sensors, each with up to a $16 \times 16$ array, can be distributed to fully cover a given area while ensuring fully coherent distributed MIMO sensing. Although several multi-antenna UWB radar systems exist~\cite{CMOS_UWB}, none provide separate RF chains for Tx and Rx antennas in active radar modes.
Motivated by the observations in Sec.~\ref{sec:motivate}, we now describe Medusa’s novel hardware design, its efficient deployment and operation, and the tightly integrated software models that fully leverage its distributed architecture.

%

\begin{figure*}[htb]
\vspace{-3ex}
\subfigure[Daughterboard]{
\centering
\includegraphics[width=0.19\textwidth]{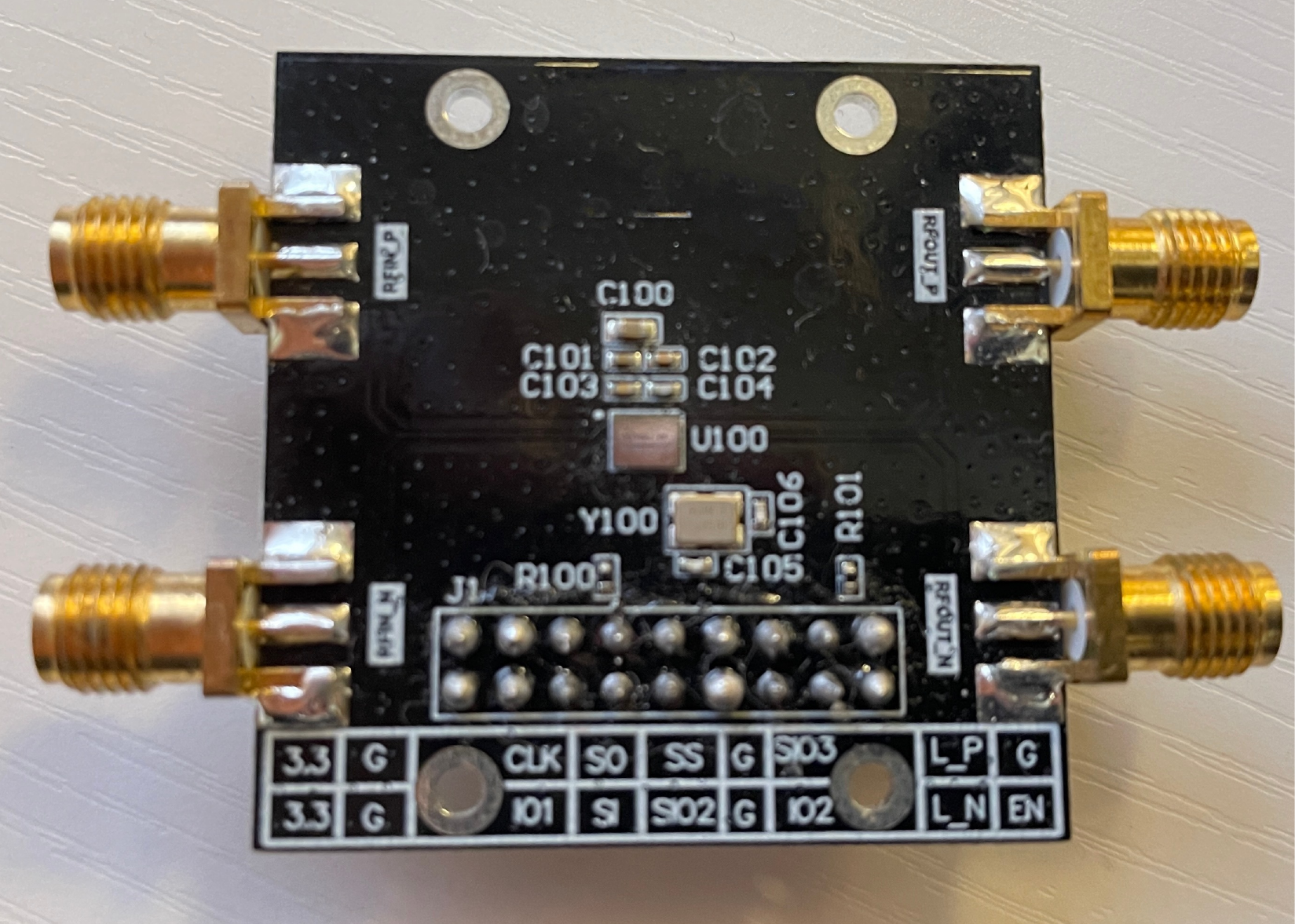}
\label{fig:daughterboard}
}
\hspace{-1.1ex}
\subfigure[Base board]{
\includegraphics[width=0.19\textwidth]{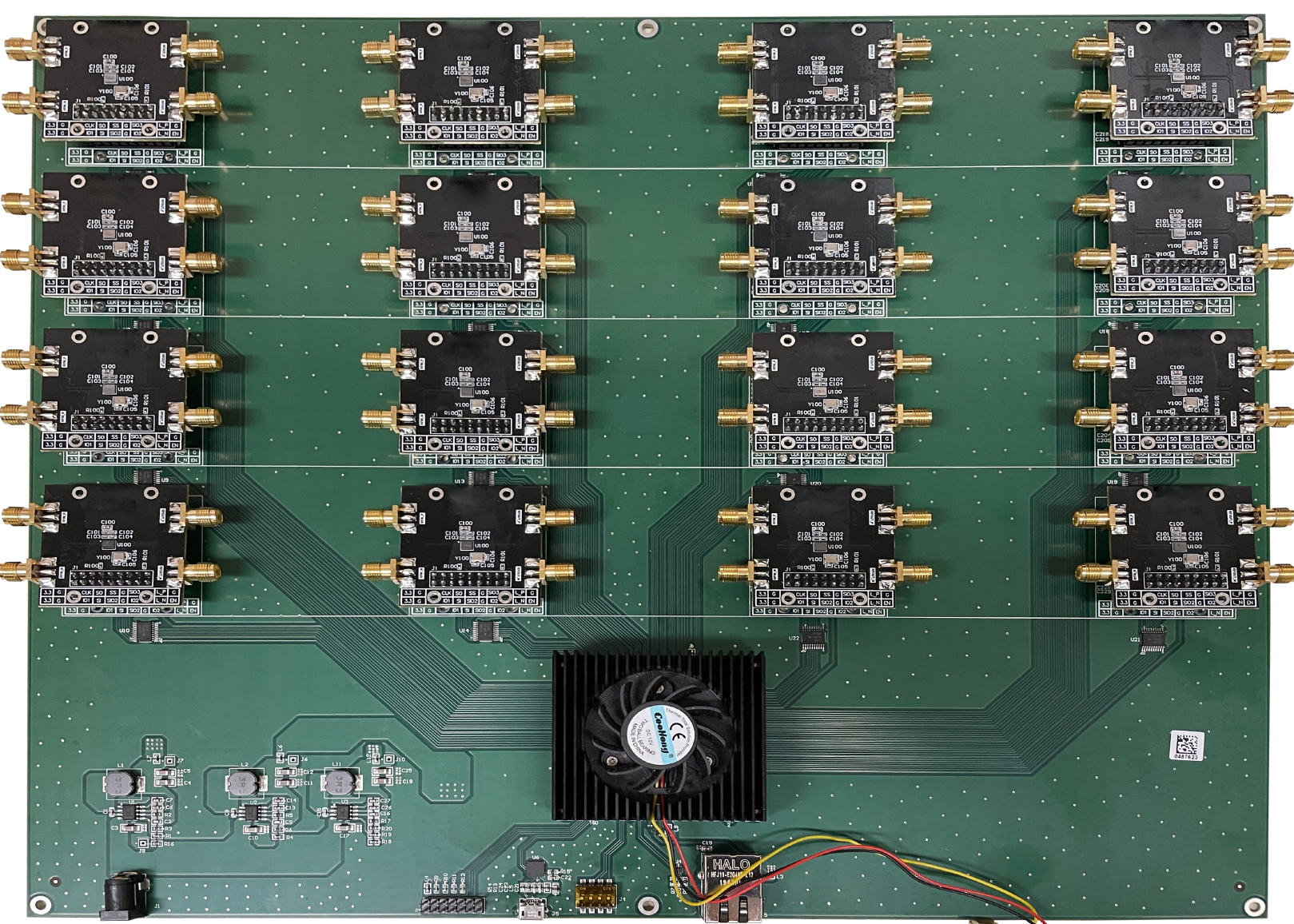}
\label{fig:motherboard}
}
\hspace{-1.1ex}
\subfigure[Connection]{
\includegraphics[width=0.18\textwidth]{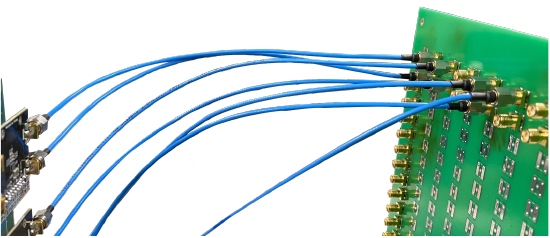}
\label{fig:antennaboard}
}
\hspace{-1.1ex}
\subfigure[Antenna board $16\times16$]{
\includegraphics[width=0.18\textwidth]{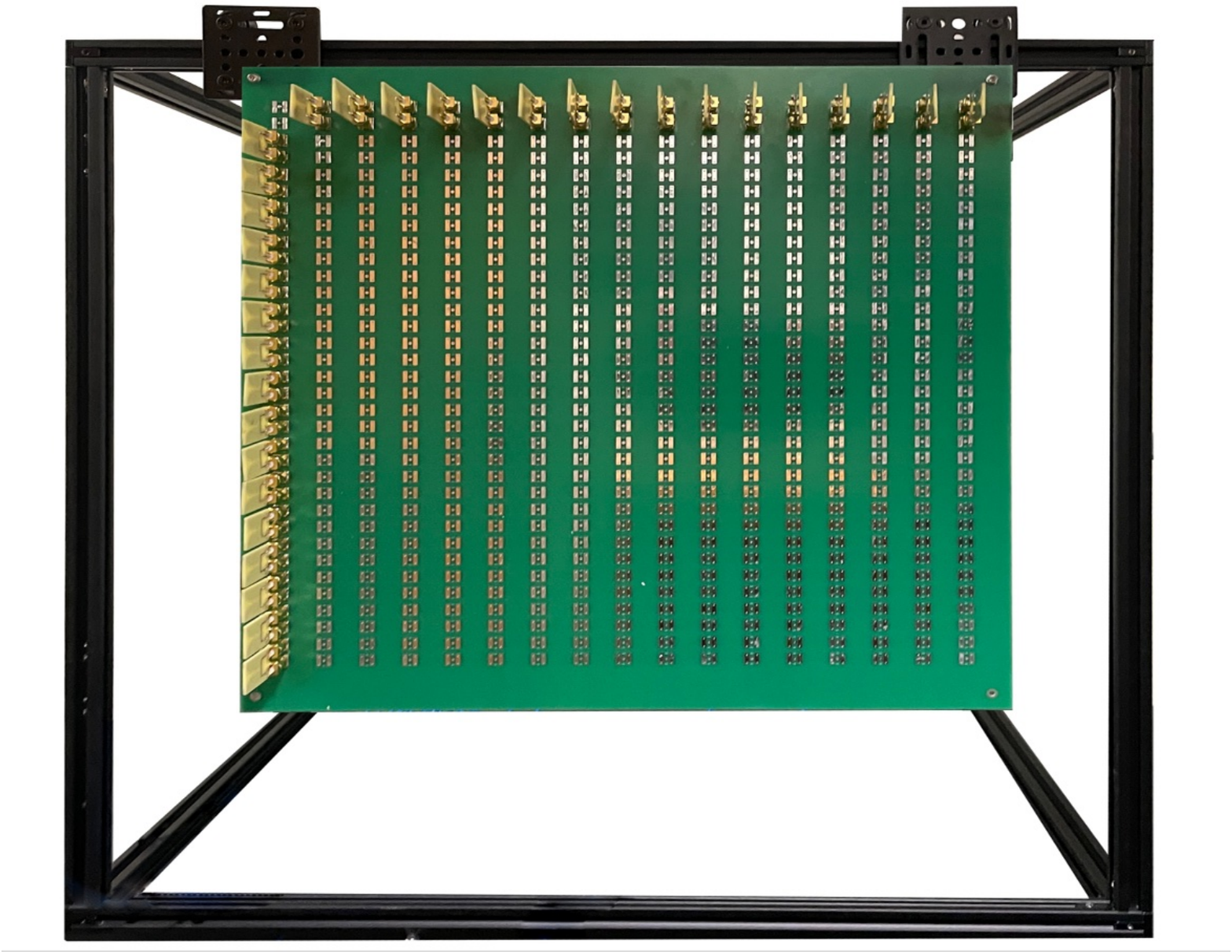}
\label{fig:antennaboard4x4}
}
\hspace{-1.1ex}
\subfigure[Antenna element]{
\includegraphics[width=0.2\textwidth]{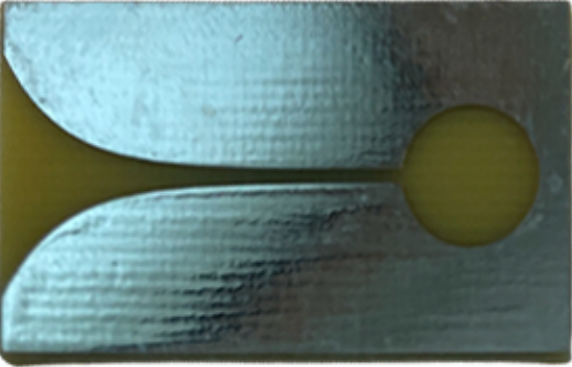}
\label{fig:antenna}
}
\caption{Medusa hardware platform.}
\label{fig:hardware}
\vspace{-2ex}
\end{figure*}

\begin{figure*}[htb]
\vspace{-0.5ex}
  \centering
  \includegraphics[width=2\columnwidth]{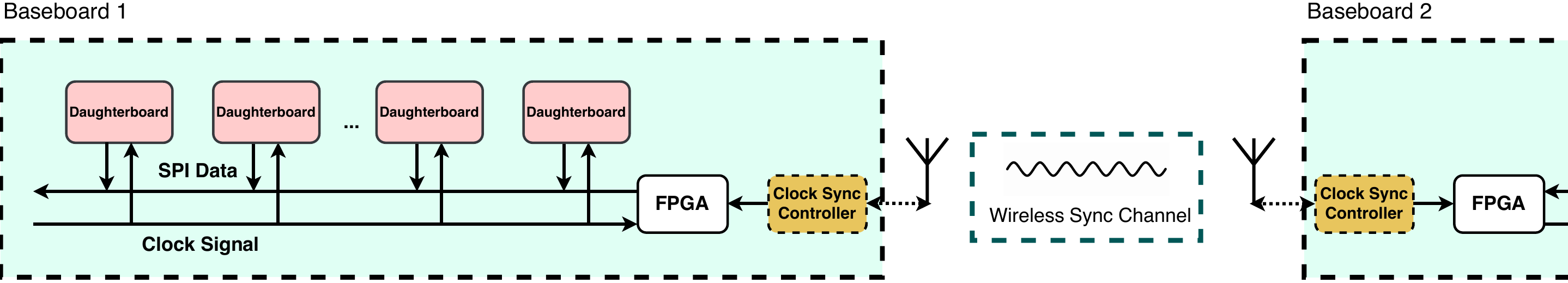}
  \vspace{-1ex}
  \caption{Medusa's scalable design with all elements clock synchronization.}
  \label{fig:coherent}
  \vspace{-1ex}
\end{figure*}

\vspace{-2ex}
\subsection{Medusa Hardware Design}
\vspace{-0.5ex}

Medusa consists of a \emph{baseboard} onto which multiple RF \emph{daughterboards} (up to 16). Each daughterboard is built around a Novelda X4 UWB radar chip~\cite{novelda} which drives one Tx and one Rx antenna. 
These antennas are mounted on a separate \emph{antenna board} that is connected to the daughterboard via RF cables. These different components of Medusa are shown in Fig.~\ref{fig:hardware}.

\vspace{-1.5ex}
\subsubsection{Medusa Radar Baseboard}
\vspace{-0.5ex}
The radar baseboard comprises a Xilinx Zynq UltraScale+ MPSoC FPGA~\cite{zynqultrascale} and socket interfaces for 16 radar daughterboards. Each socket is connected to the Zynq FPGA via 16 SPI buses and a pair of differential RF clock lines. This  design enables significant operating flexibility across the daughterboards: all 16 daughterboards to be operated as a single coherent MIMO radar; alternatively, the daughterboards can be divided into subgroups of smaller MIMO radars, with each subgroup operating independently of the others. 
Timing errors within the clock signals and SPI to the daughterboards must be eliminated to ensure coherent MIMO radar operation. 
\system distributes a phase-coherent Low-Voltage Differential Signal (LVDS) clock signal from the Zync FPGA to all attached daughterboards over the differential RF lines. 
Impedance matching is also carefully calibrated to ensure equal clock and SPI line lengths, further reducing any timing skews in the hardware.
For coherent MIMO operation, Medusa must also ensure coherence across the internal state of all X4 UWB radar chips. 
Finally, the baseboard streams I/Q data from all daughterboards, in real-time, to a host PC via a 10GbE Ethernet connection.

\vspace{-1.5ex}
\subsubsection{Medusa Daughterboard}
\vspace{-0.5ex}
Each daughterboard is built around a Novelda X4 chip~\cite{novelda}. The daughterboard routes differential clock signals and SPI commands from the baseboard to the X4 chip and forwards I/Q data from the X4 radar back to the baseboard, all with minimal time delay. Each daughterboard is physically pluggable into the baseboard via an 18-socket interface.

\vspace{-1.5ex}
\subsubsection{Medusa Antenna Design}
\vspace{-0.5ex}
Each daughterboard drives one Tx and one Rx antenna. To minimize errors in vital sign monitoring, Medusa employs custom high-gain directional Vivaldi antenna elements that provide optimal SNR for radar returns, as shown in Fig.~\ref{fig:antenna}. The Novelda X4 chip employs differential RF lines for TX and RX, requiring $100\Omega$ differential antennas. These antennas are connected to the daughterboard using SMA connectors.

\vspace{-0.5ex}
\subsubsection{Phase Offset Correction}
\vspace{-0.5ex}
The phase of the \emph{downstream} wireless clock received at each client contains an RF propagation dependent offset. To estimate this offset, each client transmits an \emph{upstream} differential clock back to the server, generated from its recovered downstream reference and therefore carrying the same propagation-dependent phase information.
By comparing the upstream clocks received from all clients, the server estimates the relative phase offsets among clients and communicates the corresponding corrections back to them. Each client then applies its correction, as shown in Fig.~\ref{fig:coherent}, ensuring phase coherence across clients. The absolute phase offset between the server and clients, however, remains unknown under this scheme. Therefore, in normal Medusa operation, the server does not participate in distributed MIMO sensing with the clients; instead, it can operate its own baseboard and daughterboards as an independent local MIMO radar.

\vspace{-1.5ex}
\subsection{Medusa Deployment - Diversity vs. SNR}
\vspace{-0.35ex}
To achieve optimal performance, it is crucial to find a tradeoff of two key factors - \textit{Diversity vs. SNR}: Diversity gains from spatially distributed MIMO arrays and SNR gains from each co-located MIMO antenna array.  We provide an analytical discussion of this trade-off, supported by empirical experiments.

\emph{\textbf{Discussion:}} Medusa shows that single-view radars capture only partial vital sign information due to signal blockages and subject movement, especially in NLoS conditions. The measurements vary significantly with the subject's orientation.
To achieve optimal performance, Medusa balances SNR and spatial diversity. Mathematically, any reflected signal \(S_t\) can be represented as a linear combination basis functions:
\[
S_t = \sum_{i=1}^{N} \alpha_i \phi_i,
\]
where \(\{\phi_i\}_{i=1}^{N}\) are components corresponding to distinct viewing angles, and $\alpha_i$ represents the amplitude of the $i_{th}$ component.
Since a complete representation of the target’s reflected signals requires decomposition into at least four orthogonal directional components, our empirical experiments demonstrate that a \textit{single-view} radar capture only partial vital sign information due to signal blockages and subject movement, particularly under NLoS conditions. In contrast, using four vantage points distributed around the subject effectively covers the full ($360^\circ$) with minimal blind spots.

\begin{figure}[htb]
    \vspace{-1ex}
     \centering
     \subfigure[1m LoS]{
         \includegraphics[width=0.45\columnwidth]{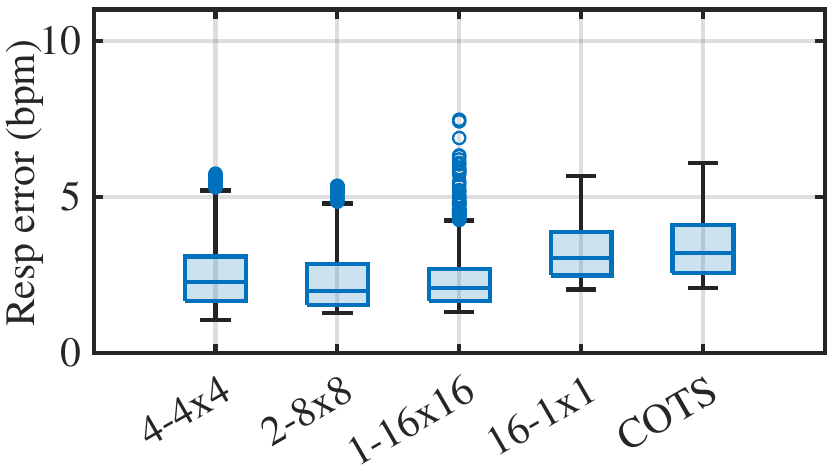}
         \label{fig:BPMCDF4x4vs16x16vs8x8vs1x1_50cm}
     }
     \subfigure[5m LoS]{
         \includegraphics[width=0.45\columnwidth]{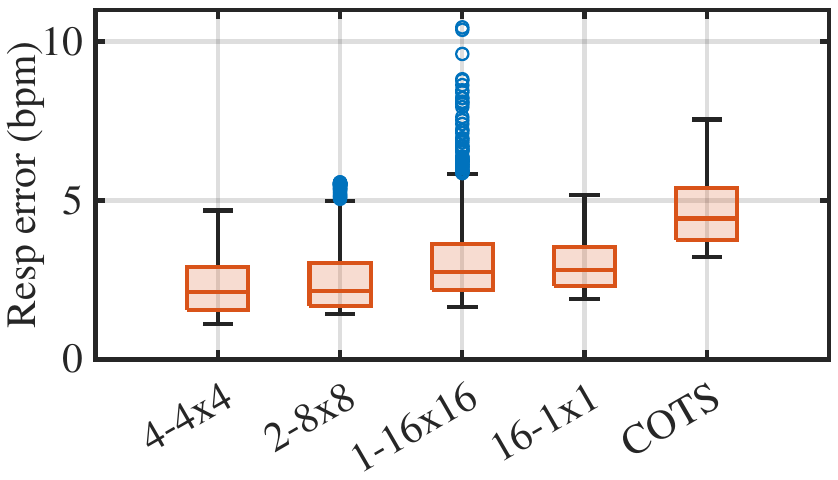}
         \label{fig:BPMCDF4x4vs16x16vs8x8vs1x1_1_5m}
     }
     \subfigure[1m NLoS]{
         \includegraphics[width=0.45\columnwidth]{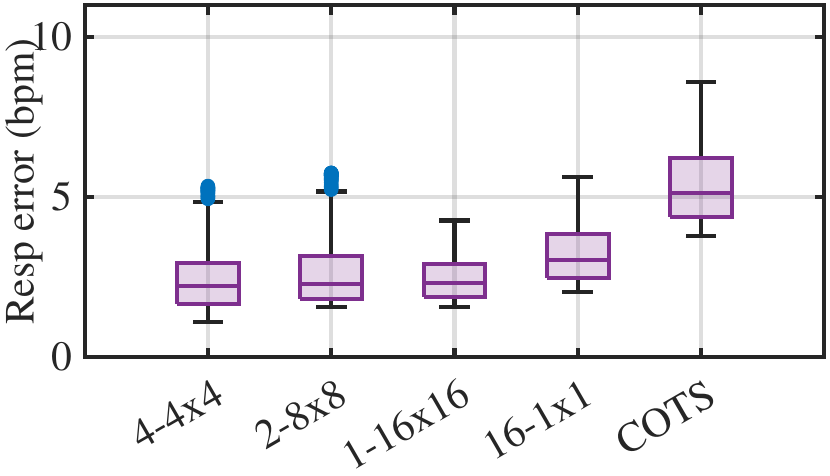}
         \label{fig:BPMCDF4x4vs16x16vs8x8vs1x1_3m}
     }
     \subfigure[5m NLoS]{
         \includegraphics[width=0.45\columnwidth]{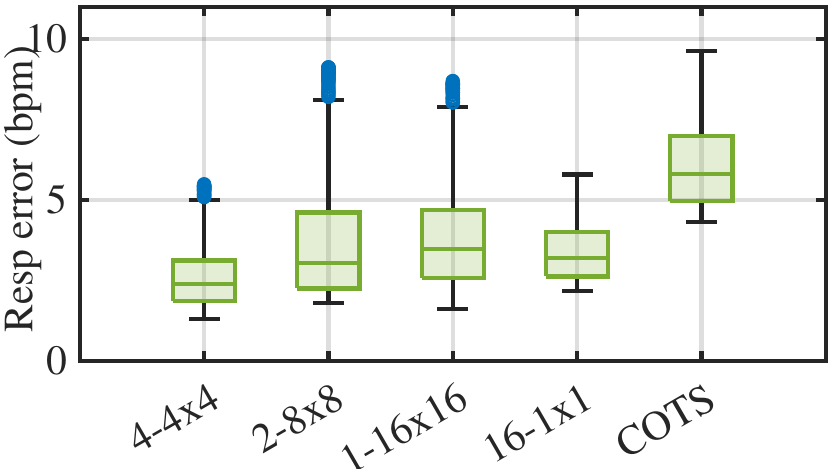}
         \label{fig:BPMCDF4x4vs16x16vs8x8vs1x1_5m}
    }
    \vspace{-2.2ex}
     \caption{Respiration rate  errors for various MIMO array configurations at 1m and 5m distances in both LoS and NLoS scenarios.}
        \label{fig:BPMCDF4x4vs16x16vs8x8vs1x1}
        \vspace{-2.1ex}
\end{figure}

\emph{\textbf{Experiments:}} Then we empirically conduct thorough experiments leveraging the Medusa's modular design to validate various configurations (e.g., one $16\times16$, two $8\times8$, four $4\times4$ and sixteen $1\times1$ radars) and evaluating the measured breaths per minute~(bpm) accuracy with a static target positioned at different indoor locations (see Fig~\ref{fig:top_view_deploy}) for \textbf{both LoS and NLoS scenarios}. (Details on ground truth and data collection can be found in Sec~\ref{sec:eval}).
Fig~\ref{fig:BPMCDF4x4vs16x16vs8x8vs1x1} shows a subset of our experimental results. 
In LOS conditions, when the target is close to a radar (e.g., 1m), the $16\times16$ configuration delivers the best performance due to its high SNR gains with an average respiratory error of 2.01 bpm. However, when the target moves away to 5m, the $16\times16$ configuration experiences a decline in accuracy as the SNR decreases, resulting in mean respiratory error of 2.98 bpm. Meanwhile, the spatially distributed four $4\times4$ and two $8\times8$ configurations together help compensate for the SNR loss at individual radars. 
In NLOS scenarios, 
the 16 $1\times1$ distributed single antenna arrays, while benefiting from spatial distribution and achieving more stable performance than single-view $16\times16$ MIMO arrays, do not gain any SNR advantages, resulting in a median BPM error of 3.41 bpm at 1m and 4.12 bpm at 5m.
The four $4\times4$ configuration, on the other hand, offers the best trade-off (median BPM error 2.11 bpm) between SNR and diversity gains and perform the best. 
Consequently, Medusa adopts the four $4\times4$ configuration.

\begin{figure}[htb]
\vspace{-0.5ex}
  \centering
  \includegraphics[width=0.6\columnwidth]{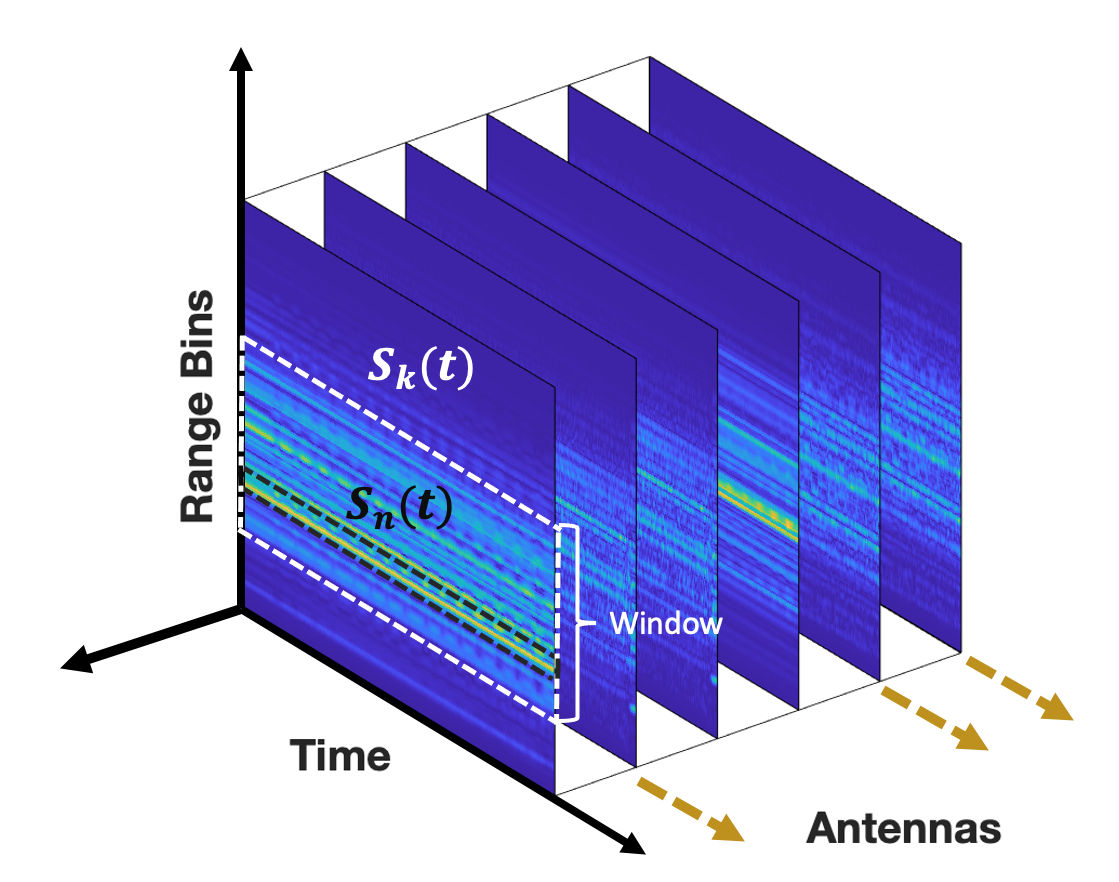}
  \vspace{-1ex}
  \caption{Radar frame. The radar data structure of Medusa comprises CIR matrices associated with each radar element corresponding to each antenna.}
  \label{fig:rangebins}
  \vspace{-2ex}
\end{figure}

\vspace{-1.5ex}
\subsection{Human Vitals Sensing with Medusa}
\vspace{-0.5ex}
The X4 chip leverages UWB time-of-flight pulses to measure distance, producing 186-bin range measurements per pulse, with each bin capturing the amplitude and phase of reflected signals. Human activities and vital signs (e.g., breathing, heartbeats) induce variations in these reflections, which mix with environmental motion and multipath distortions. Medusa employs Time Division Multiplexing (TDM), sequentially transmitting pulses from a single TX while all 16 radar receivers coherently capture the returns. Each receiver generates 186 range bins, representing reflection amplitudes at different distances. A data frame consists of radar returns from a single transmission cycle, and a single probe aggregates 16 sequential frames, corresponding to 16 radar pulse transmissions. These I/Q data streams are transmitted over a 10 GbE link and processed by an unsupervised ML model to extract human vital signs while filtering out environmental noise. Figure~\ref{fig:rangebins} illustrates the collected range data over time.

\begin{table*}[htb]
\vspace{-0.5ex}
\centering
\begin{tabular}{|l|c|c|c|}
\hline
                           & \textbf{Medusa Platform}                 & \textbf{IWR1443BOOST}          & \textbf{AWR2243 Cascade}      \\ \hline
Frequency Band             & 6.5-9.5GHz                      & 76-81 GHz                      & 76-81 GHz                     \\ \hline
TX/RX                      & 16TX/16RX                       & 3TX/4RX                        & 12TX/16RX                     \\ \hline
Azimuth Array              & 256 element virtual array       & 12 element virtual array       & 86 element virtual array      \\ \hline
Max Angular Resolution & 0.448 degree & 9.53 degree & 1.4 degree \\ \hline
Min Spacing Separation & 0.039m at 5m                    & 0.841m at 5m                   & 0.122m at 5m                  \\ \hline
Frame Rate (FPS) & 50 -- 200                   & 10                   & 5                  \\ \hline
\end{tabular}
\caption{Comparison of Medusa platform, TI IWR1443BOOST, and TI AWR2243 Cascade.}
\label{table:UWBmmWave}
\vspace{-3ex}
\end{table*}

\subsubsection{Extracting Human Vitals}
Radar data captures a blend of breathing and heartbeat signals, along with unwanted motion and multipath distortions, making separation challenging, especially for multiple individuals with unknown activity. While \textbf{FFT} and peak-finding suffice in controlled settings, they often fail in complex real-world environments.

Medusa decomposes the human vital signals from the raw data mixed with other interfering signals using independent component analysis (ICA)~\cite{hyvarinen2013ica}. Consider a \emph{single} radar receiver, $k\in \{1,\ldots,M\}$, where $M$ is the total number of radar receivers. Let $X_k(t) =[x_1(t),\ldots,x_N(t)]^T$ be a vector of the $N$ source signals induced by human respiration, heartbeats, and motion for one or more monitored individuals. These signals are combined in a non-linear fashion at the radar sensors as
\begin{equation} \label{eqn1}
\vspace{-0.5ex}
S_k(t) = f([x_1(t),\ldots,x_N(t)]^T) = [s_1(t),\ldots,s_N(t)]^T
\vspace{-0.5ex}
\end{equation}

\begin{figure}[htb]
\vspace{-0.5ex}
  \centering
  \includegraphics[width=1.0\columnwidth]{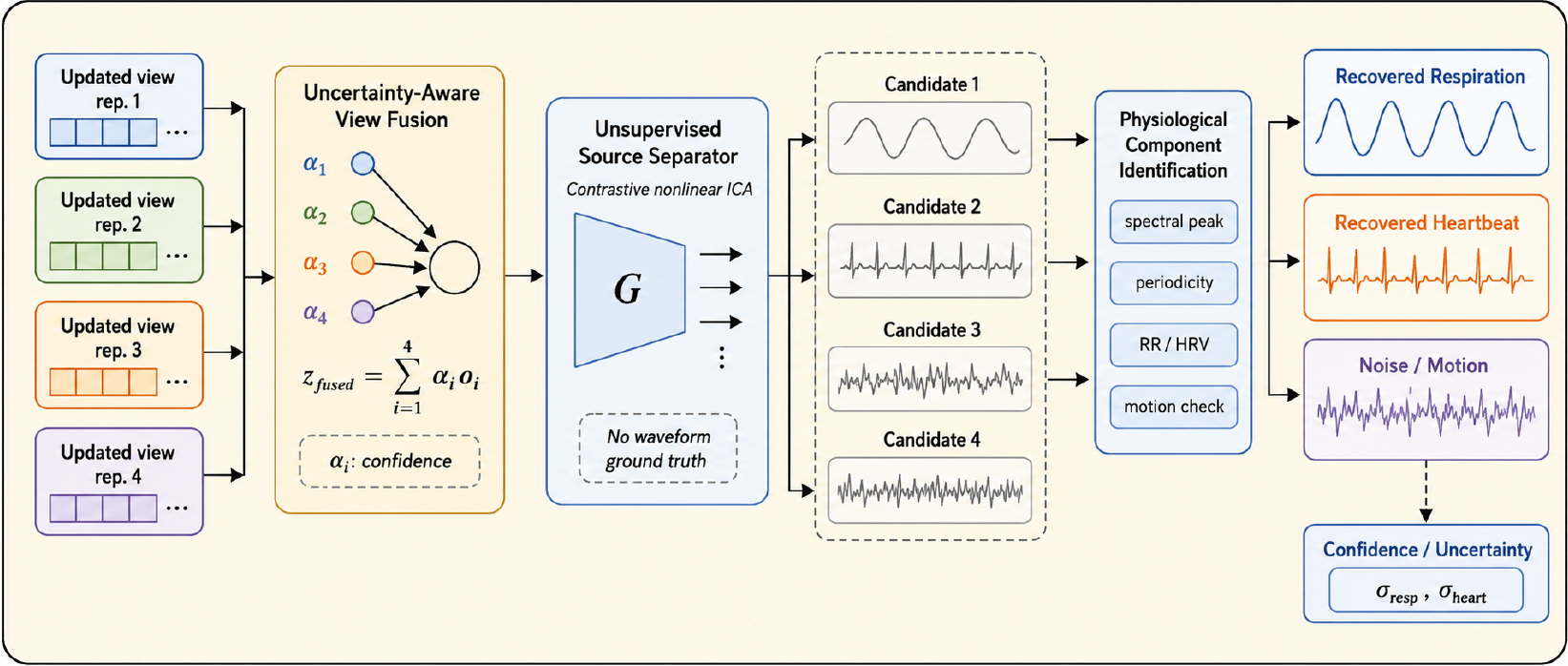}
  \vspace{-1.8ex}
  \caption{Architecture of our unsupervised learning model to decompose and recover the waveforms.}
  \label{fig:modelworkflow}
  \vspace{-2ex}
\end{figure}
\noindent where $S_k(t)$ is the vector of $N$ received signals at the $k^{th}$ radar receiver. In Medusa, these $N$ signals $s_n(t)$ arrive in $N$ different range bins. 
The objective of non-linear ICA is to find the approximate inverse $f^{-1}$ to recover the source signals $X_k(t)$ from $S_k(t)$. To do this, we employ a contrastive learning model, as shown in Fig.~\ref{fig:modelworkflow}.

In Medusa, $S_k(t)$ represents a window of $N$ range bins from receiver $k$ at some time $t$, where $1 \leq N \leq 186$. We generate positive $Y_k(t)$ and negative $Y_k^*(t)$ augmented samples from $S_k(t)$ for contrastive training, defined as
\begin{equation}
\vspace{-0.5ex}
    Y_k(t) = \left(\begin{array}{c}S_k(t)\\S_k(t-T)\end{array}\right), \; 
    Y_k^*(t) = \left(\begin{array}{c}S_k(t)\\S_k(t-\delta)\end{array}\right)
    \vspace{-0.5ex}
\end{equation}
where $T$ is a constant and $\delta$ is a randomly selected time offset. 

Let $\mathbf{E}(\cdot)$ define the encoder network used in the contrastive model. If we train the model to discriminate between $\mathbf{E}(Y_k(t))$ and $\mathbf{E}(Y_k^*(t))$, we obtain the representation of $h(S_k(t)) \approx f^{-1}(S_k(t))$~\cite{Aapo2016}. 
Note however, that this model does not yet account for coherent radar signals from multiple receivers. To this end, Medusa uses a multi-head attention step~\cite{NIPS2017_attention} in its unsupervised model as described next. By leveraging nonlinear ICA, the model effectively decomposes mixed signal vectors. Additionally, heart rate and respiration rate exhibit distinct characteristics in the frequency domain, enabling their separation.

\subsubsection{Multi-Receiver Fusion}
Instead of utilizing the received signals from each antenna $S_k(t)$ directly, Medusa uses a \textit{multi-head attention layer} to fuse signal information from all the radar receivers, separately for the positive and negative augmented signals, prior to the contrastive training:
\begin{align}
\vspace{-0.5ex}
        [Z_1(t),\ldots,Z_W(t)] &= \mathbf{A}([Y_1(t),\ldots,Y_M(t)]),\nonumber\\
        [Z_1^*(t),\ldots,Z_W^*(t)] &= \mathbf{A}([Y_1^*(t),\ldots,Y_M^*(t)]).        
\end{align}
Here $W$ is the number of heads in the attention layer~\cite{NIPS2017_attention}, and $\mathbf{A(\cdot)}$ is the attention layer function that maps the $M$ radar receiver signals into $W$ head outputs. We then train the contrastive model to discriminate between the encoded pair $\mathbf{E}(Z_w(t))$ and $\mathbf{E}(Z_w^*(t))$ of each head output $~w \in \{1,\ldots,W\}$. 


%
\subsubsection{Features and Vital Signs Identification}
\begin{figure}[htb]
\vspace{-1ex}
  \centering
  \includegraphics[width=0.9\columnwidth]{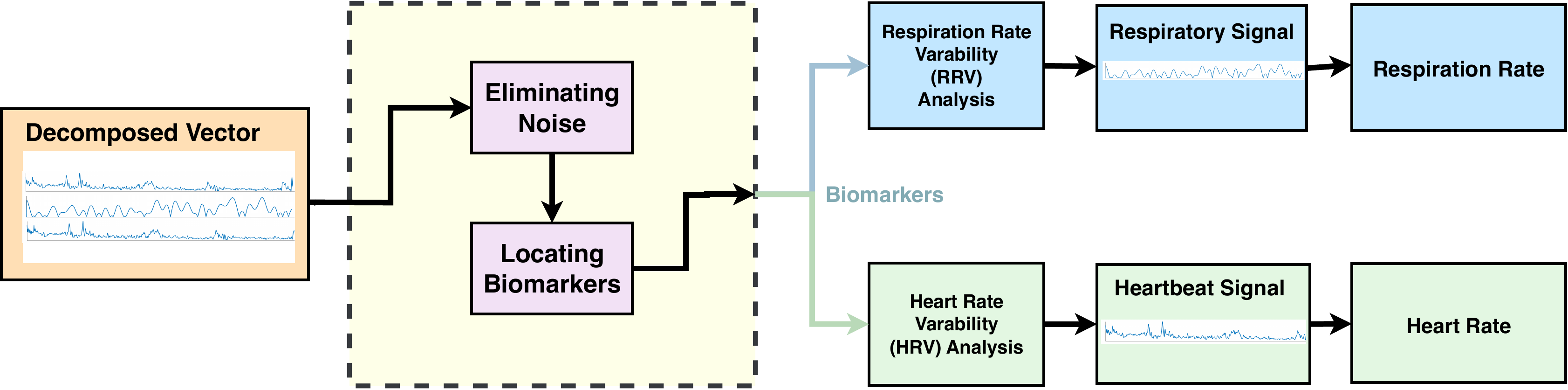}
  \vspace{-1ex}
  \caption{Workflow of separating respiration, heart-rate, and motion patterns from the mixed radar signals. The resulting heartbeat and breath waveforms are identified.}
  \label{fig:distinguishworkflow}
  \vspace{-2ex}
\end{figure}

Medusa identifies the breathing and heartbeat signals by analyzing the Respiratory Rate Variability (RRV)~\cite{Makowski2021neurokit} and Heart Rate Variability (HRV)~\cite{neuralkit2HRV} of each of the $W$ signals, as illustrated in Fig.~\ref{fig:distinguishworkflow}. RRV and HRV are key indicators of general health and respiratory or cardiac complications. Normal breathing and heart-rate exhibit relatively constant rates and volumes, but variations within these rhythms are labeled as RRV and HRV, respectively. We use RRV and HRV analysis to identify the correct breathing or heart-beat signals from the output features of the trained model. 
We use the extracted waveforms to identify if they are in respiratory rhythm or show normal variations in heart rate, and distinguish them into respiratory waveforms, heart-rate signals, or noise. This allows us to detect and track vital signs in the radar data accurately.


\vspace{-1ex}
\subsubsection{Window Selection}
The UWB radar X4 chip extracts 186 range bins representing reflection amplitudes in distance. To reduce computing time, we use a window of size $N$ to select a range bin subset before inputting the data into the model. We select the $N$ consecutive bins with the highest reflected signal power, as these are most likely to contain human reflections. We choose a 30-bin section with higher amplitudes, reducing computational time for model training. This approach enables more effective detection and tracking of human reflections in radar data.

\vspace{-0.5ex}
\section{Implementation and Evaluation}
\label{sec:eval}
\vspace{-0.5ex}
We conduct \system's evaluation in two parts. 
First, we compare \system's custom-designed platform accuracy with that of the COTS mmWave MIMO radar(as indicated in Table~\ref{table:UWBmmWave}) used previously 
in~\cite{movifi,morefi,uwbinfocom2022,mobi2sense,octopus,V2iFi,gong_UbiComp2021}. Next, we 
evaluate \system's efficacy to measure the respiration and heart rate of multiple diverse targets in real-time, in
real-world environments.

\vspace{-1.5ex}
\subsection{\textbf{\system} MIMO radar Micro-Benchmark}

\begin{figure}[!htb]
\vspace{-1.5ex}
\centering
\begin{minipage}[t]{.5\linewidth}
    \vspace{-0.5ex}
    \includegraphics[width=0.95\columnwidth]{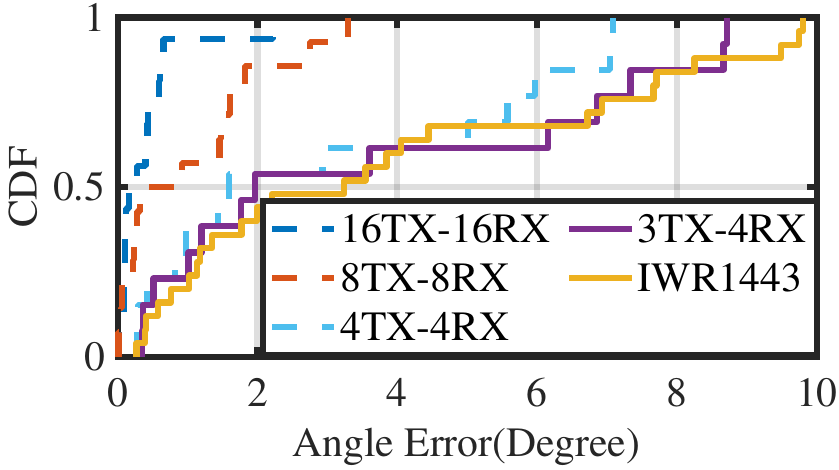}
    \vspace{-1ex}
    \caption{AoA accuracy.}
    \label{fig:angleestimation}
    \vspace{-2ex}
\end{minipage}%
\begin{minipage}[t]{.5\linewidth}
    \vspace{-0.5ex}
    \includegraphics[width=0.95\columnwidth]{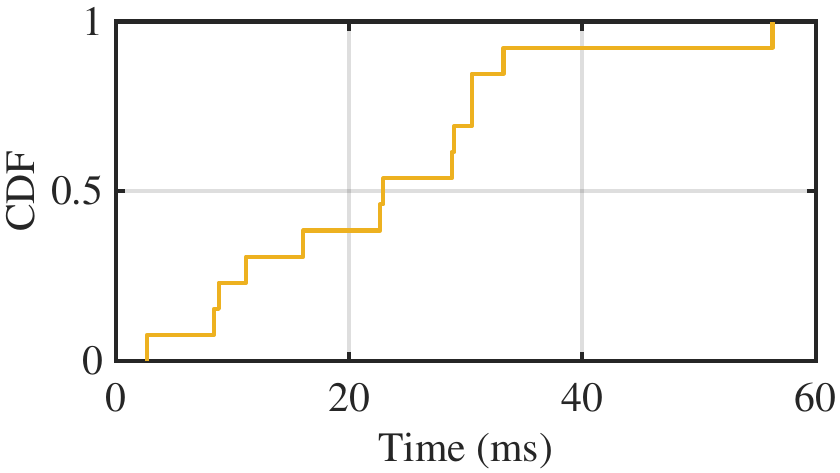}
    \vspace{-1ex}
    \caption{CDF of Latency.}
    \label{fig:CDFLatency}
    \vspace{-2ex}
\end{minipage}%
\end{figure}

\emph{\textbf{AOA accuracy}}: We compare the AoA accuracy of the \system's MIMO radar with the COTS UWB and mmWave radars in Table~\ref{table:UWBmmWave}. We
use a reflector box and position it at various distances and angles (but at the same height as the radar) measure the range and AoA of the strongest reflected signal, and compare it with the ground-truth.
Fig.~\ref{fig:angleestimation} shows the AoA performance. For AoA in LoS, measured at 5m distance, \system MIMO radars
perform better than the COTS mmWave radar. Median angular errors for \system is only 2 deg and a max AoA error of 8.2 deg, while the COTS mmWave radar's 
AoA median errors are 6.2 deg, while max angular error can go up to 12.5 deg.

\begin{figure}[ht]
\vspace{-1.5ex}
     \centering
    \subfigure[Carrier frequency offset (CFO)]{
        \includegraphics[width=0.45\columnwidth]{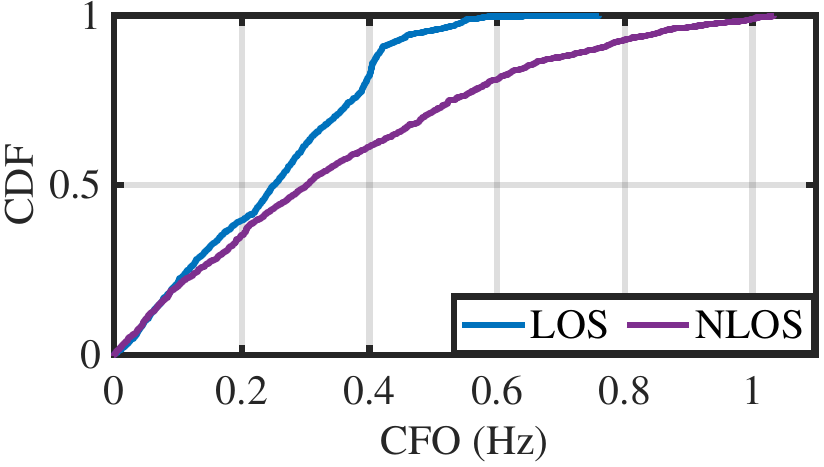}
        \vspace{-1ex}
        \label{fig:clockCFO}
     }
     \subfigure[Phase offset]{
        \includegraphics[width=0.45\columnwidth]{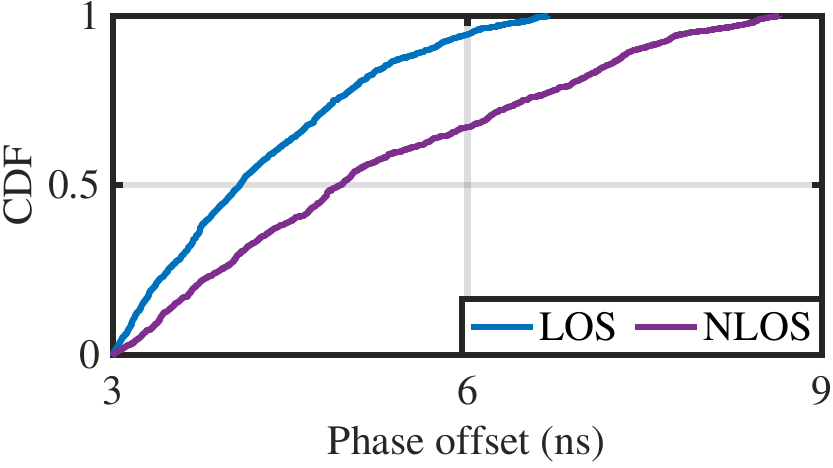}
        \vspace{-1ex}
        \label{fig:clockphaseoffset}
     }
     \vspace{-1ex}
     \caption{Wireless clock sync. in LoS and NLoS} 
    \label{fig:evalsyncwireless}
     \vspace{-2ex}
\end{figure}

\noindent \emph{\textbf{System Latency and Wireless Coherency:}} Fig.\ref{fig:CDFLatency} shows a median latency of 22.94ms for respiration waveform reconstruction, from the time radar data 
was received, proving that \system can indeed run in real-time.

Next, we show the efficacy of \system's of wireless clock synchronization within reasonable ranges. As shown in Fig.~\ref{fig:clockCFO}, the median carrier frequency offset (CFO) is 0.25Hz in LOS and 0.3Hz in NLOS. Fig.~\ref{fig:clockphaseoffset} displays the phase offset during the wireless synchronization of clock signals. 

\vspace{-1.5ex}
\subsection{Experiment Setup and Baseline}
\label{sec:vital_exp}
\begin{figure}[htb]
\vspace{0.5ex}
 \centering
  \includegraphics[width=0.9\columnwidth]{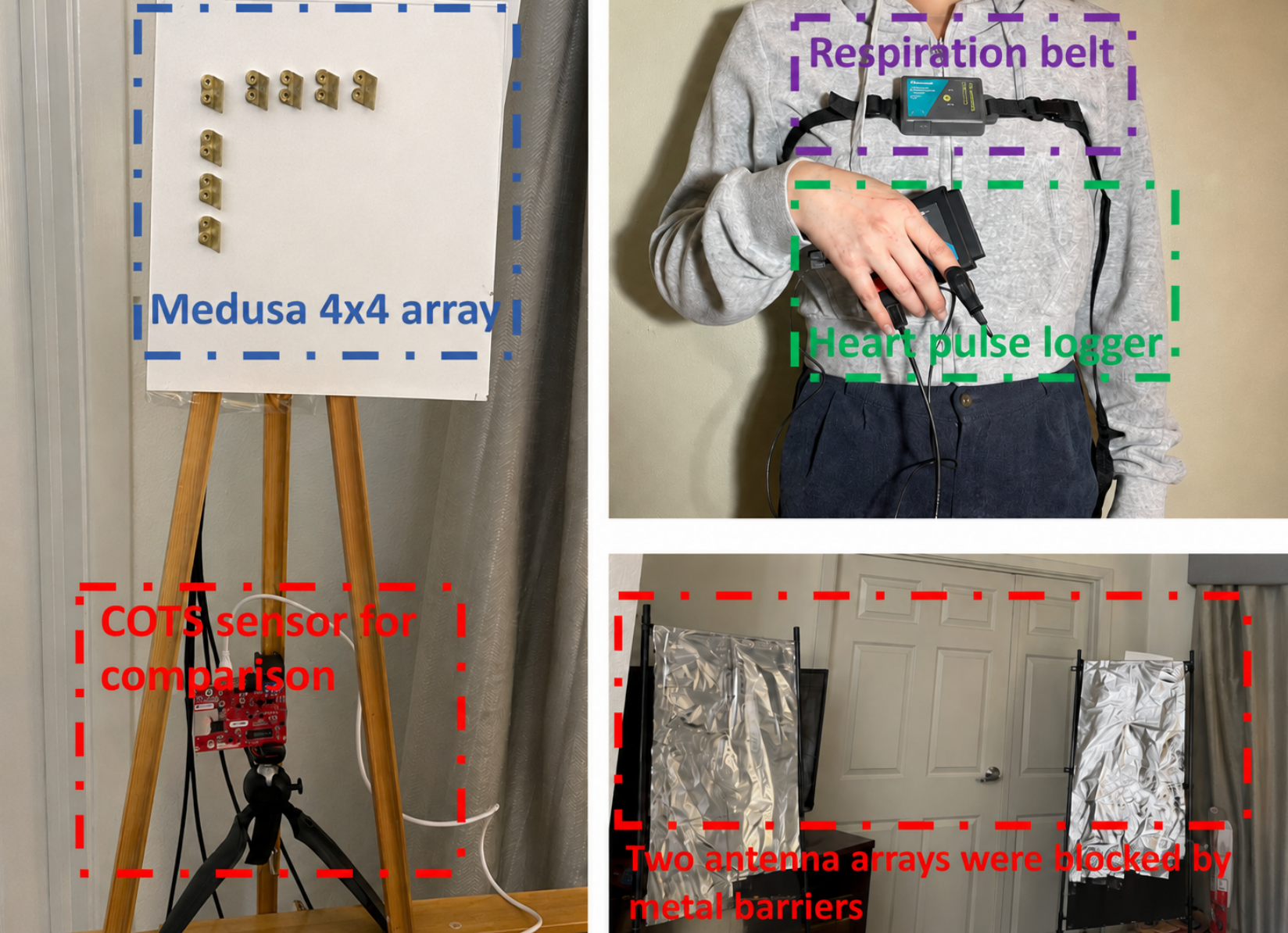}
  \vspace{-1ex}
  \caption{Experiment setup with MIMO radars, GT respiration belt, heart rate sensor and IWR1443BOOST mmWave sensor. And also the screen shield barrier for NLOS experiments. (Photos were taken in a bedroom, capturing only the setup, yet accurately reflecting a real-world scenario.)}
  \label{fig:deploymentEval}
  \vspace{-2ex}
\end{figure}

\emph{\textbf{Ground truth (GT):}} 
We utilized Vernier's breathing belt \cite{respirationsensor} and heart rate sensor \cite{heartsensor} as Ground Truth (GT) for vital signs, commonly used in prior research \cite{morefi,movifi,gong_UbiComp2021,RFVital_UbiComp21,adib:2015}. All participants wore these sensors during experiments, with necessary consent and IRB approvals obtained beforehand.  
Figure~\ref{fig:deploymentEval} illustrates the experiment setup. 
The TI radar sensor was directed at the chest, while the UWB radar, operating at a lower frequency with a wider beam, covered the entire body, allowing flexible positioning. The respiration belt and radar base board were network-connected via different interfaces: the radar used Ethernet, while the belt transmitted data via a Bluetooth hub connected over a Wi-Fi router.

\noindent \emph{\textbf{Baselines:}}
In our evaluation, we employ two baselines. For respiration pattern detection, we utilize prior work~\cite{morefi,movifi,gong_UbiComp2021} that uses Novelda UWB radar as a baseline. For heart rate monitoring, we compare \system with RFSCG~\cite{ha2020contactless}, which is implemented on the TI mmWave radar IWR1443~\cite{TImmWave1443}.

%

\noindent \emph{\textbf{Data Collection:}}
We collected data from 27 volunteers (14 men, 13 women) aged 21-34 years (average age 25), weighing 52-102 kg (average 81.2 kg), and ranging in height from 164-187 cm (average 175 cm). Data collection took place in four indoor locations, primarily in a lab measuring $18ft \times 30ft$ (540 sq ft), which exceed the size of average US bedrooms. Volunteers wore casual attire during 10-minute sessions, and none reported cardiovascular issues.
We collect data of each person performing the following actions:  (1) \emph{static dataset}: standing, sitting with their body oriented in different directions. 
(2)\emph{mobile dataset:} Arbitrary walking and jogging across a room in different directions, standing up and sitting down in continuous and staggered motion, 
jogging at the same spot with different body orientations, and performing various hand gestures. In total, we collected 3.75TB of data: 1.89TB from static activities and 1.86TB from mobile activities. We split the data into training and test sets with an 80:20 ratio, allocating approximately 3.04TB for training and 710GB for testing. The model underwent training with the designated training dataset and evaluation with the test dataset.

\vspace{-1.5ex}
\subsection{Unsupervised model}
\vspace{-0.5ex}
\system leverages an unsupervised learning model to recover human vital signals, and here we describe the network architecture and training process in detail.

\noindent \emph{\textbf{Network structure:}} The unsupervised multi-head attention model includes a classifier and an encoder with a multi-head attention layer, implemented using TensorFlow by Python 3.8. Binary cross-entropy is utilized to classify loss. We added a multi-head attention layer before the Dense layer to assign weights to each MIMO antenna array, which helps the system focus on the most significant input data from influential antennas.

\begin{figure}[htb]
\vspace{-3ex}
     \centering
    \subfigure[University Lab]{
        \includegraphics[width=0.45\columnwidth]{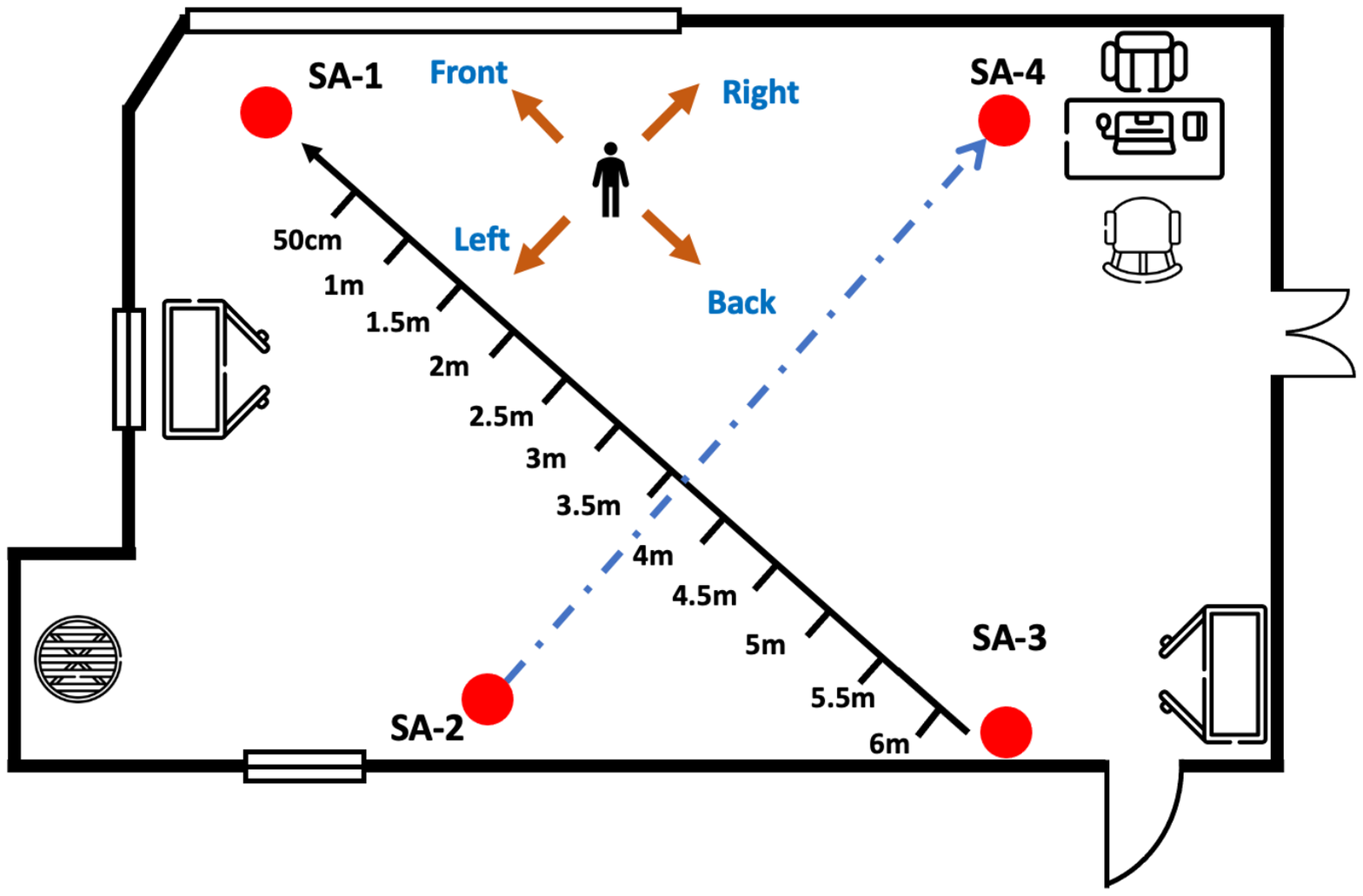}
        \vspace{-1ex}
        \label{fig:top_view_lab}
     }
     \subfigure[Bedroom]{
        \includegraphics[width=0.4\columnwidth]{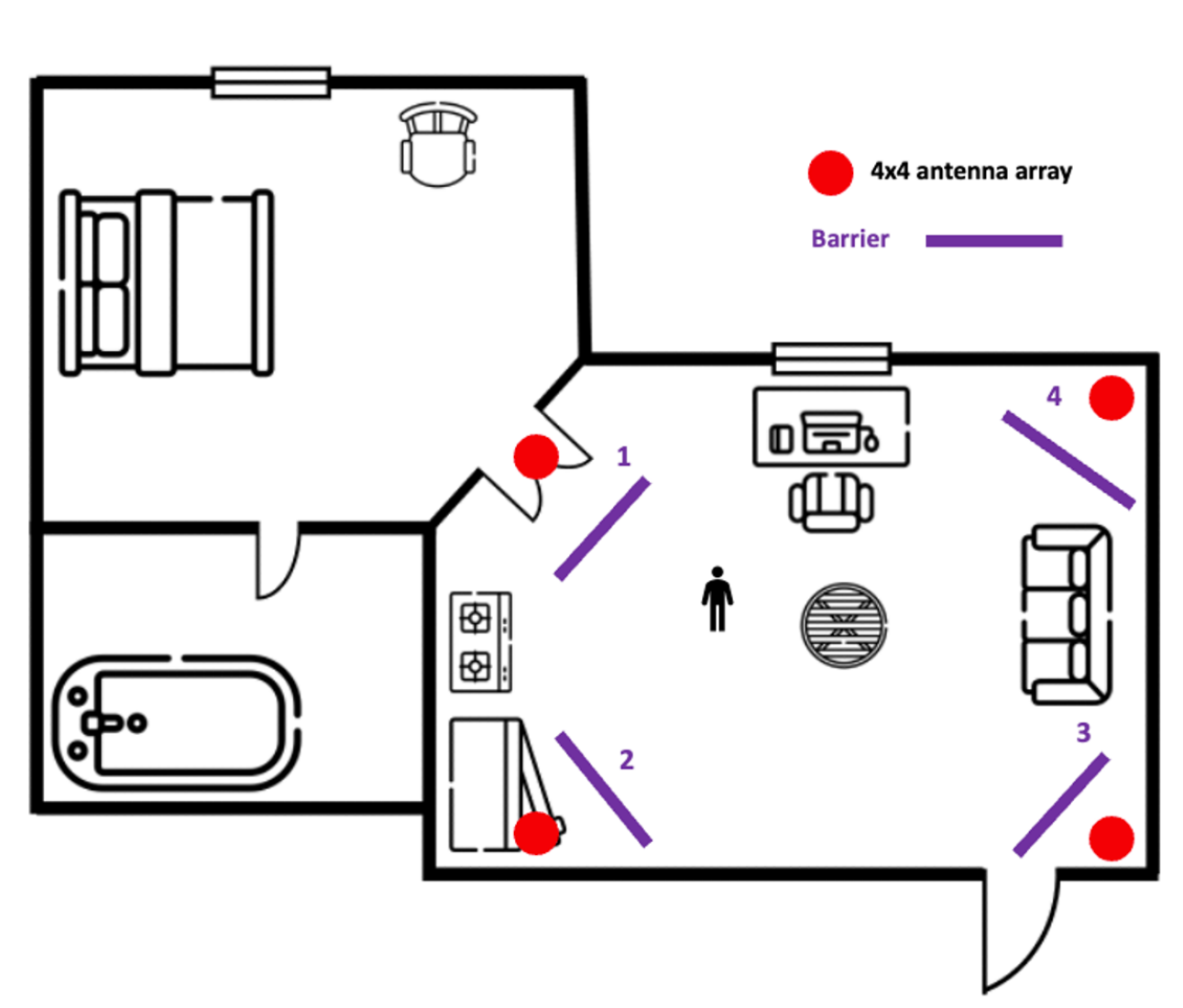}
        \vspace{-1ex}
        \label{fig:top_view_room}
     }
     \vspace{-1ex}
  \caption{Top-down view of the experimental setup, showing four 4×4 radar placements (highlighted in red) across different environments~(lab and bedroom).}
  \label{fig:top_view_deploy}
  \vspace{-3ex}
\end{figure}

\noindent \emph{\textbf{Training details:}} The model is trained on a PC featuring an AMD Ryzen 6900 CPU and an RTX 3080 Ti graphics card. Upon deploying the trained model, data is transmitted from the baseboard to the PC through a high-speed Ethernet interface. For a measurement duration of 2 minutes, the collected data has a size of 1.8GB and requires approximately 2 hours of model training using a single RTX 3080 Ti graphics card with 200 iterations. During training, datasets are split 80:20 for training and testing, using the leave-one-person-out cross-validation (LOPO CV) technique. The model is trained with the training dataset and evaluated with the test dataset, iteratively executed for each individual, ensuring each person's data is used as the test dataset once. This process is carried out iteratively for each participant in the dataset, guaranteeing that each individual's data is used as the test dataset precisely once. 

\subsection{Performance Evaluation:}
\label{sec:vital_eval}
\subsubsection{\system: Accuracy and Robustness}
\label{sec:trained_eval}
We begin evaluating \system's robustness to monitor the Respiration-Per-Minute (BPM) of static and mobile
users in LoS and NLoS when they are oriented in different directions. The experiment setup with the room layout, radar locations, items of furniture, as well
as distances between radars are shown in Fig~\ref{fig:top_view_deploy}. The four $4\times4$ subarray of \system is located at corners and we named them \textbf{SA-1}, \textbf{SA-2}, \textbf{SA-3} and \textbf{SA-4}.  
NLoS experimental setup employs a common approach in that we use screen shields as obstacles between the target and radars rather than placing the target near the radars. This configuration more accurately simulates real-world NLoS conditions.

\begin{figure}[htb]
\vspace{-1.5ex}
    \centering
    \subfigure[Respiration waveform reconstructed for a moving target in LoS]{
      \includegraphics[width=0.9\columnwidth]{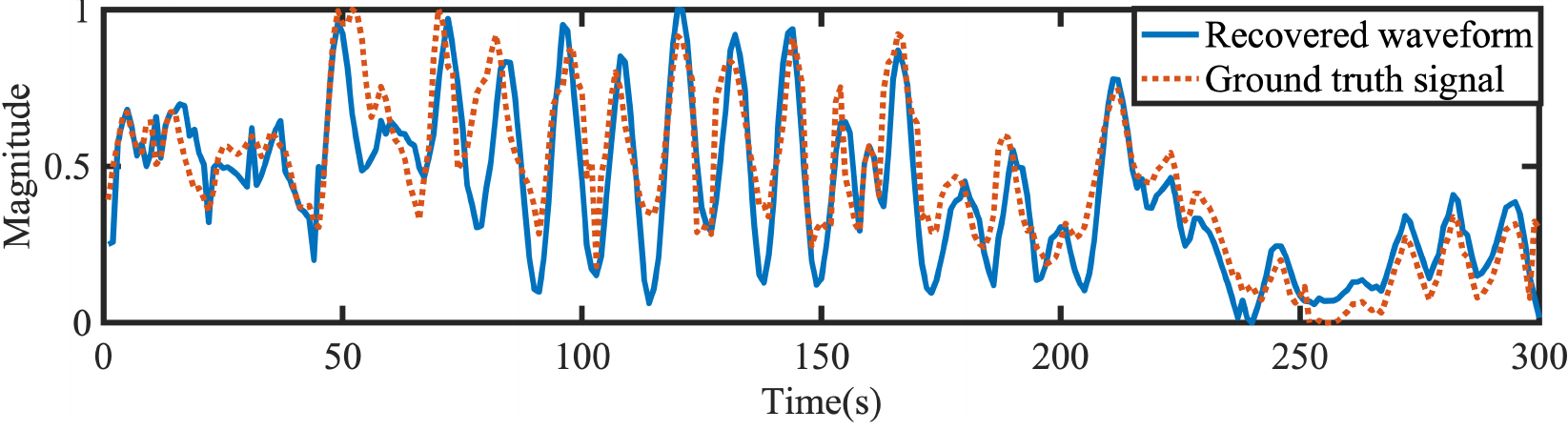}
      \vspace{-1ex}
      \label{fig:waveform_1}
    }
    \vspace{-1.5ex}
    \subfigure[Respiration waveform reconstructed for a static target in NLoS]{
        \includegraphics[width=0.9\columnwidth]{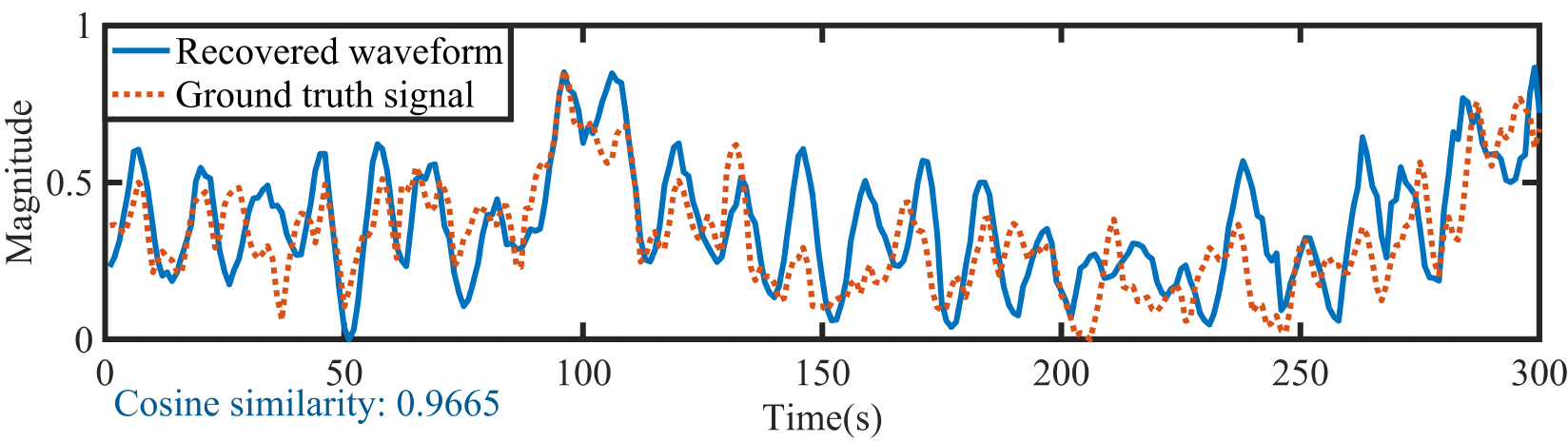}
        \vspace{-1ex}
        \label{fig:waveform_2}
    }
    \vspace{-2ex}
    \caption{Respiration waveforms reconstructed by \system .} 
    \label{fig:waveform}
    \vspace{-2ex}
\end{figure}

\begin{figure}[htb]
    \vspace{-2ex}
     \centering
     \subfigure[Front]{
       \includegraphics[width=0.45\columnwidth]{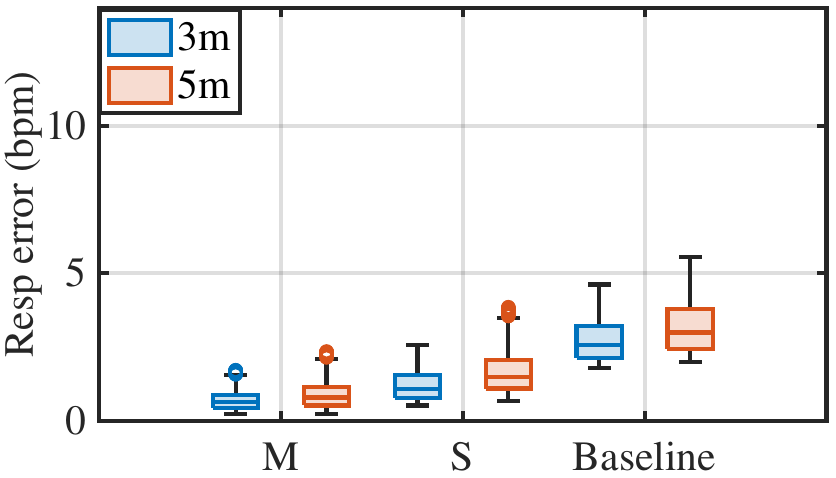}
       \label{fig:BPM3mfront}
     }
     \subfigure[Left and Right]{
        \includegraphics[width=0.45\columnwidth]{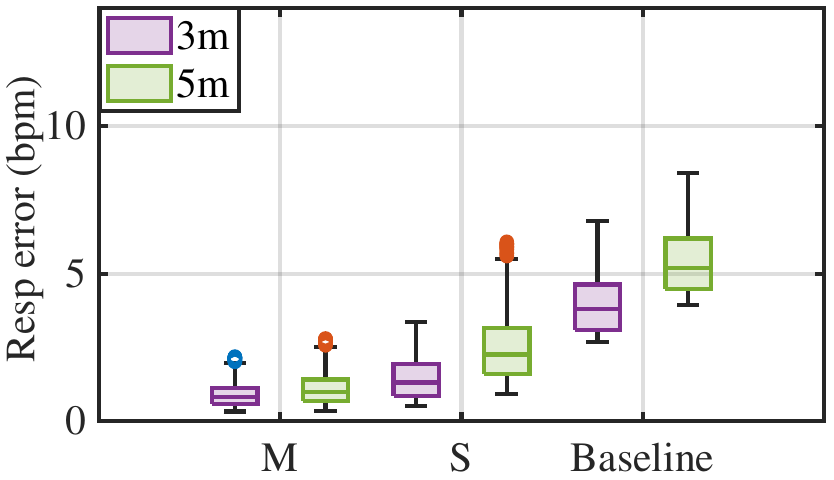}
        \label{fig:BPM3mleft}
     }
     \vspace{-2ex}
     \caption{LoS Static targets: BPM with different orientations. Comparison of \system~(`M'), Co-located single-view MIMO radar~(`S'), and Baseline.} 
    \label{fig:BPM3m}
    \vspace{-2ex}
\end{figure}

\emph{\textbf{Waveform Recovery:}}
Fig~\ref{fig:waveform} displays the reconstructed waveform from \system for a randomly chosen participant positioned at 5 meters and facing left. The reconstructed waveform closely 
resembles the actual ground-truth waveform, exhibiting a cosine similarity of 0.987.
Fig~\ref{fig:waveform_2} demonstrates the increased complexity of respiration waveform recovery in Non-Line-of-Sight (NLoS) conditions, as evidenced by the reduced cosine similarity of 0.966.

\emph{\textbf{Static Line-of-Sight~(LOS):}} 
Fig~\ref{fig:BPM3m} presents the box-plot of respiration accuracy for static users (measured individually) when they are either standing or sitting and facing two different directions (body orientations) at distances 3m and 5m distance. 
The respiratory rate accuracy decreases at 5m for the single-view $16 \times 16$ radar facing frontward, dropping by 12.1\% when compared to the accuracy when facing left.
The baseline's performance is worse than the single-view $16 \times 16$ radar due to its poor SNR, especially when the user is 5m away, facing either front or left. \system, on the contrary, with $4\times4$ MIMO distributed subarrays compensates for the drop in SNR (see Fig~\ref{fig:deploymentEval}) when the target is 5m away by combining signals from the other three radars, providing an improvement of 5.8\% and 18.5\% over the baseline and single-view deployments, respectively.
The benefits of \system are seen further when the target is oriented to the left and right at the 5m mark. When the baseline and the single-view solutions suffer, \system still estimates respiration with median errors < 2.8 bpm.

%

\begin{figure}[htb]
\vspace{-1ex}
     \centering
     \subfigure[Back]{
       \includegraphics[width=0.45\columnwidth]{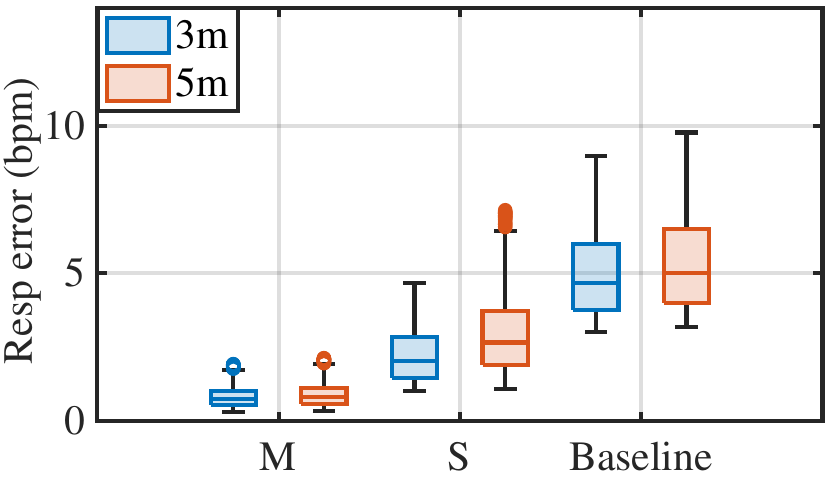}
       \vspace{-0.5ex}
       \label{fig:BPM3mfrontNLOS}
     }
     \subfigure[Right]{
        \includegraphics[width=0.45\columnwidth]{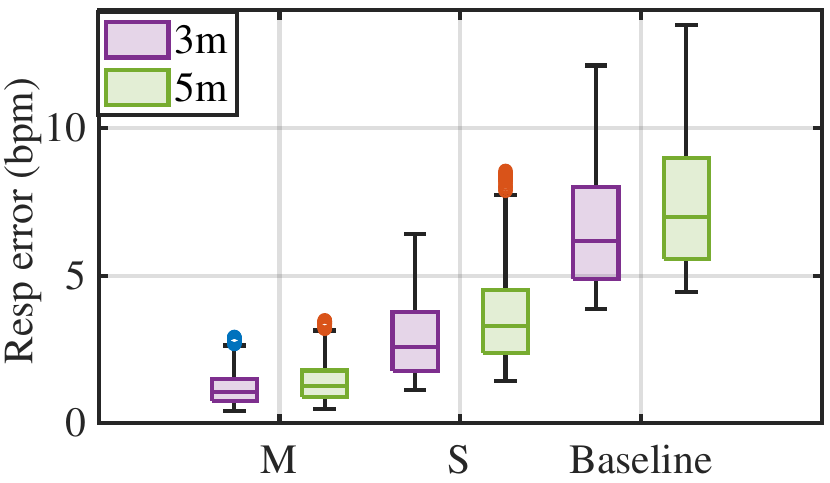}
        \vspace{-0.5ex}
        \label{fig:BPM3mleftNLOS}
     }
     \vspace{-1ex}
     \caption{NLoS static users: Respiration errors (bpm) with different orientations.} 
    \label{fig:BPM3mNLOS}
    \vspace{-2ex}
\end{figure}

\vspace{-0.5ex}
\begin{figure}[htb]
\vspace{-0.5ex}
     \centering
      \subfigure[Jogging and Walking in LoS]{
       \includegraphics[width=0.45\columnwidth]{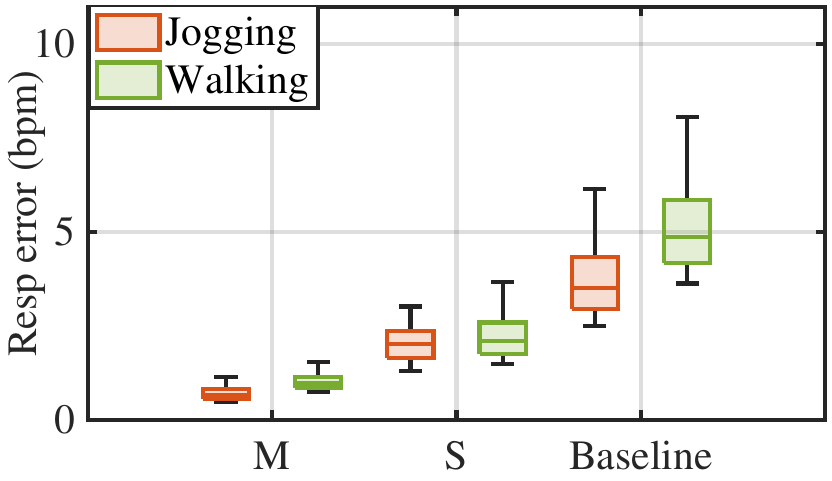}
       \vspace{-0.5ex}
       \label{fig:BPMmobileknown_NLOS_static}
     }
     \subfigure[Jogging and Walking in NLoS]{
        \includegraphics[width=0.45\columnwidth]{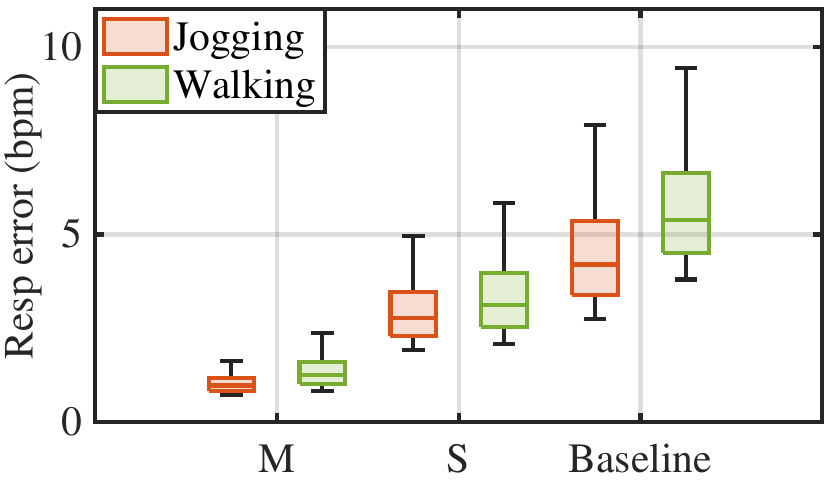}
        \vspace{-0.5ex}
        \label{fig:BPMmobileknown_NLOS_motion}
     }
     \vspace{-1ex}
    \caption{Respiration errors (bpm) for targets with movements in LoS and NLoS.} 
    \label{fig:BPMmobileknown_NLOS}
    \vspace{-2ex}
\end{figure}

\emph{\textbf{Static None-Line-of-Sight~(NLOS):}} 
To create NLoS 
conditions, we manually place the barrier screen (refer to Fig~\ref{fig:deploymentEval}) in front of the back and right radars.  
Fig~\ref{fig:BPM3mNLOS} shows the box-plot for respiration errors for all the individuals standing at the same 3m and 5m mark but
oriented back and right 
(refer to Fig~\ref{fig:top_view_lab} and Fig.~\ref{fig:top_view_room}). 
When participants face the back array, \system's median  errors are 2.15 bpm at 3 meters and 2.33 bpm at 5 meters. Meanwhile, when facing the right, the median errors measured by \system are 2.21 bpm at 3 meters and 2.47 bpm at 5 meters, a gain of 17.5\% and 21.9\% over single-view and baseline solutions at 5 meters, respectively.

\emph{\textbf{Mobility in LoS and NLoS:}} We next show the result when targets are mobile in LoS and NLoS (2 covered radars) conditions.  
Fig~\ref{fig:BPMmobileknown_NLOS} showed the box plot of the average respiration errors per target when walking and jogging in random directions in LoS and
in NLoS. In LoS, \system's measured errors are 2.15 bpm and 2.36 bpm during walking and jogging, respectively, while in NLoS, \system's measured median respiration errors are 2.67 bpm and 2.56 bpm, respectively. On the other hand, the single-view ($16 \times 16$) single radar and the baseline solutions suffer, especially in NLoS conditions, with their respective median BPM errors increasing to 14.9\% and 28\%, respectively. 
\system delivers median gains of 10.3\% and 21.7\%, and max. gains of 21\% and 38\% over the single-view and baseline solutions, respectively.

\vspace{-0.5ex}
\subsubsection{Impact on Attention-weights in NLoS.}
\label{sec:attention}
\begin{figure}[!htb]
    \vspace{-2ex}
     \centering
     \subfigure[SA-1 in NLoS]{
       \includegraphics[width=0.45\columnwidth]{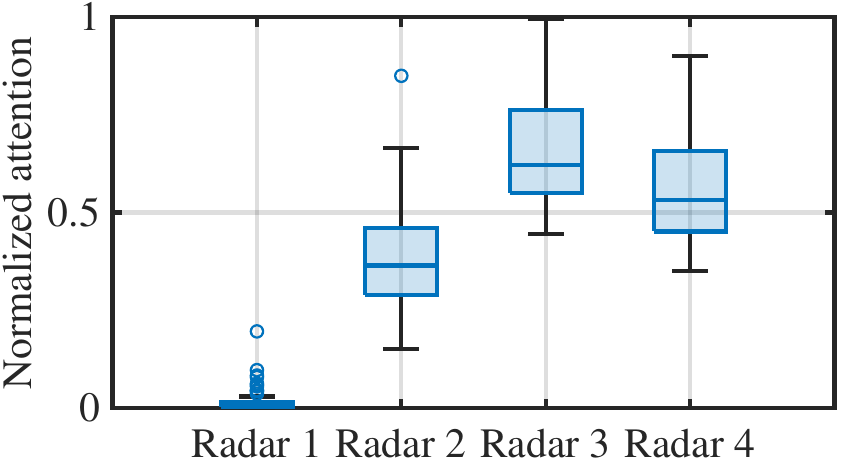}
       \vspace{-0.5ex}
       \label{fig:attentionweightspositions1}
     }
      \subfigure[SA-2 in NLoS]{
        \includegraphics[width=0.45\columnwidth]{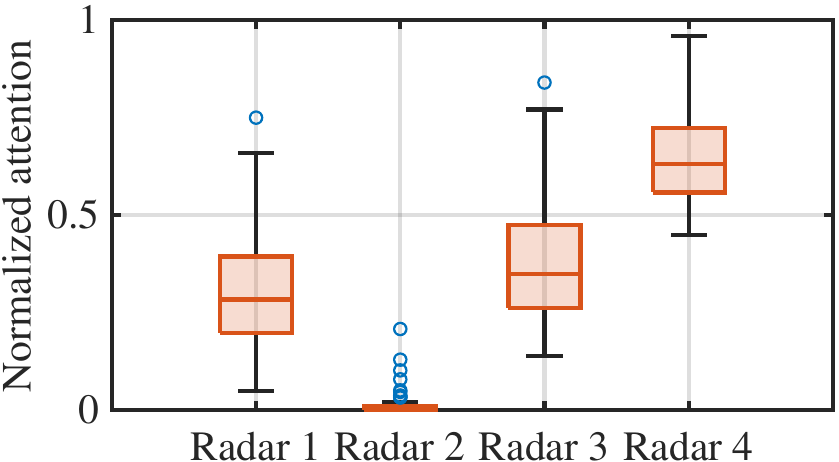}
        \vspace{-0.5ex}
        \label{fig:attentionweightspositions2}
     }
      \subfigure[SA-3 in NLoS]{
       \includegraphics[width=0.45\columnwidth]{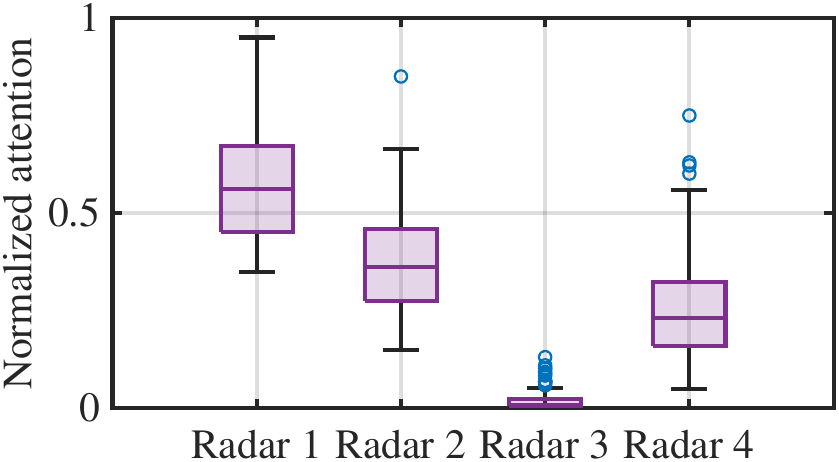}
       \vspace{-0.5ex}
       \label{fig:attentionweightspositions3}
     }
      \subfigure[SA-4 in NLoS]{
        \includegraphics[width=0.45\columnwidth]{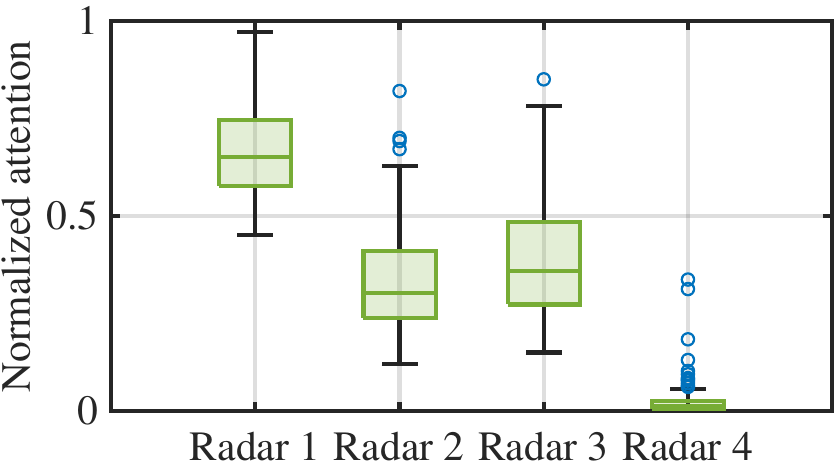}
        \vspace{-0.5ex}
        \label{fig:attentionweightspositions4}
     }
     \vspace{-1ex}
     \caption{Normalized attention weights when one of the $4\times4$ subarray (SA) of \system in NLoS.} 
    \label{fig:attentionweightspositions}
    \vspace{-2.5ex}
\end{figure}

\vspace{-0.2ex}
We show how \system modifies attention weights for each radar depending on the received signal quality. In this experiment, we sequentially block one subarray (SA) at a time (SA-1 to SA-4) and measure the respiration of static targets at 3m. Fig~\ref{fig:attentionweightspositions} illustrates the normalized attention weight distribution when different SAs are obstructed. \system reduces attention weights for radars with poor SNR (NLoS) and increases weights for radars with higher SNRs (LoS). Low SNR SAs are given lower preference, focusing instead on radars with better-received signals.

\vspace{-2ex}
\subsubsection{\system in unknown environments}
\label{sec:untrained_eval}
\vspace{-0.2ex}
Next, we evaluate \system's ability to generalize to different environments than where it was trained to ensure a true ``in-the-wild'' deployment. We train \system using data from a residence and use the model in a university lab.

\begin{figure}[htb]
    \vspace{-2ex}
     \centering
    \subfigure[Static (SS)]{
       \includegraphics[width=0.45\columnwidth]{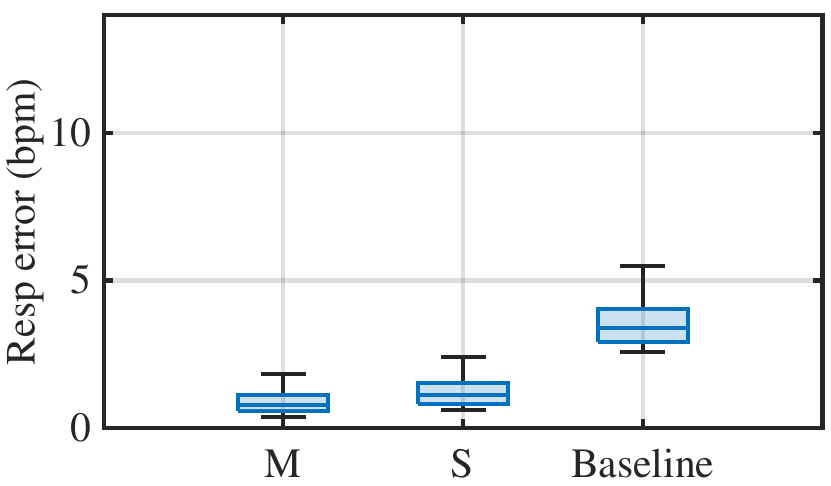}
       \vspace{-1ex}
       \label{fig:BPMmobileknown_unknown_static}
     }
     \subfigure[Jogging and walking]{
        \includegraphics[width=0.45\columnwidth]{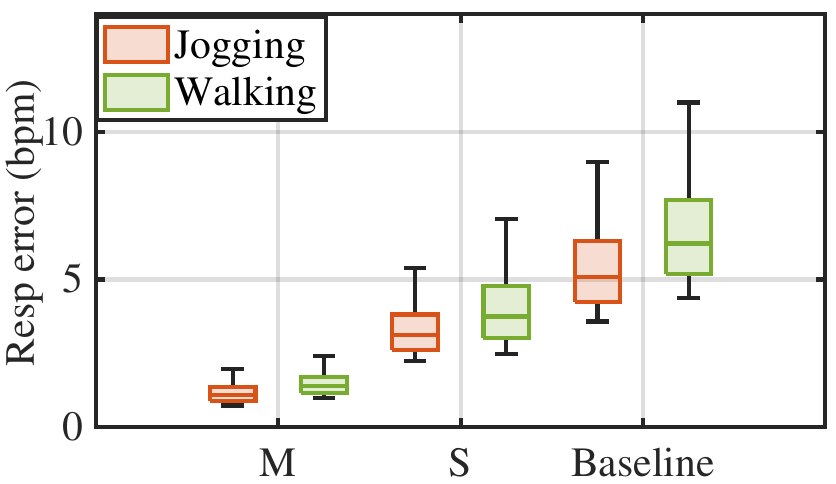}
        \vspace{-1ex}
        \label{fig:BPMmobileknown_unknown_motion}
     }
     \vspace{-2ex}
     \caption{Respiration errors (bpm) in LoS of Static and Moving targets in the untrained environment.} 
    \label{fig:BPMmobileunknown}
    \vspace{-2ex}
\end{figure}

\emph{\textbf{Targets in LoS:}} 
We perform static and mobility experiments with various targets in the radars' LoS. In the static experiment, targets stand anywhere in the room, in any orientation. In the mobility experiment, targets continuously walk and jog within the room. Fig~\ref{fig:BPMmobileunknown} displays BPM errors for static and mobile experiments. \system's median BPM errors are 1.98 bpm for static targets and 2.12 bpm and 2.46 bpm for jogging and walking, respectively. \system outperforms the single-view and baseline solutions with accuracy gains of 11.2\% and 26\%, respectively.

\begin{figure}[htb]
    \vspace{-2.2ex}
     \centering
      \subfigure[Static (SS)]{
       \includegraphics[width=0.45\columnwidth]{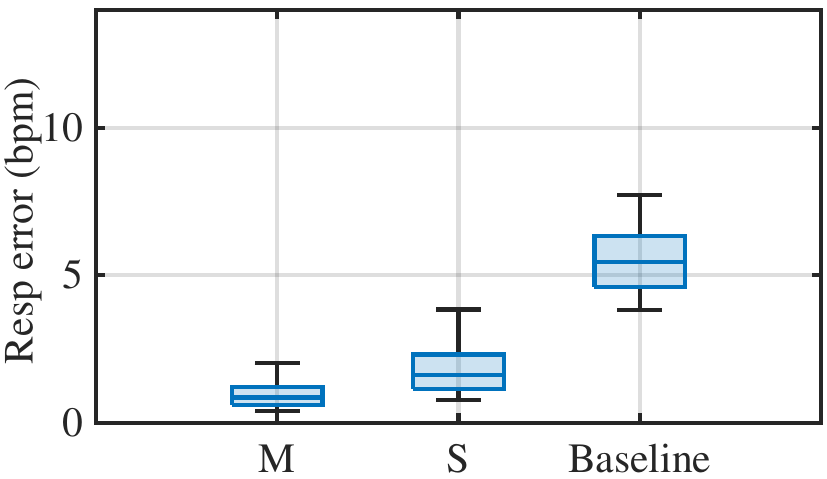}
       \vspace{-1ex}
       \label{fig:BPMmobileunknownNLOS_static}
     }
      \subfigure[Jogging and walking]{
        \includegraphics[width=0.45\columnwidth]{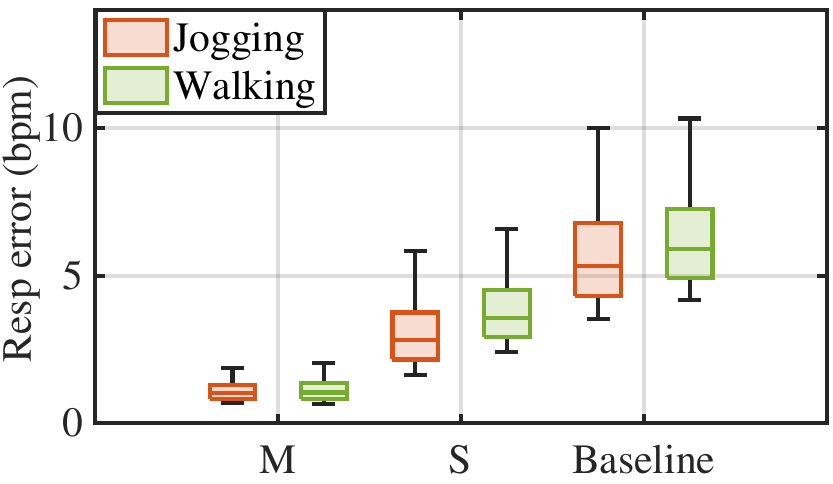}
        \vspace{-1ex}
        \label{fig:BPMmobileunknownNLOS_motion}
     }
     \vspace{-1.5ex}
     \caption{Respiration errors (bpm) in NLoS of Static and Moving targets in the untrained environment.} 
    \label{fig:BPMmobileunknownNLOS}
    \vspace{-3ex}
\end{figure}

\emph{\textbf{Targets in NLoS:}} 
We evaluate BPM errors for static and mobile targets in NLoS conditions, as shown in Fig~\ref{fig:BPMmobileunknownNLOS}. With two blocked radars, single-view and baseline solutions' median errors shoot up to 2.67 bpm and 5.98 bpm for static targets, with max errors of 18.3\% and 38\%, respectively, while \system's median error is 2.21 bpm, improving median accuracy by 2.4\% and 20.8\% over the other solutions. For mobile targets, \system accuracy gains are 2.32 bpm and 19.2\% for jogging and 2.43 bpm and 24.4\% for walking, over the single-view and baseline solutions, respectively.

\begin{figure}[htb]
\vspace{-2ex}
  \centering
    \subfigure[Cosine similarity of the recovered respiration waveform]{
       \includegraphics[width=0.45\columnwidth]{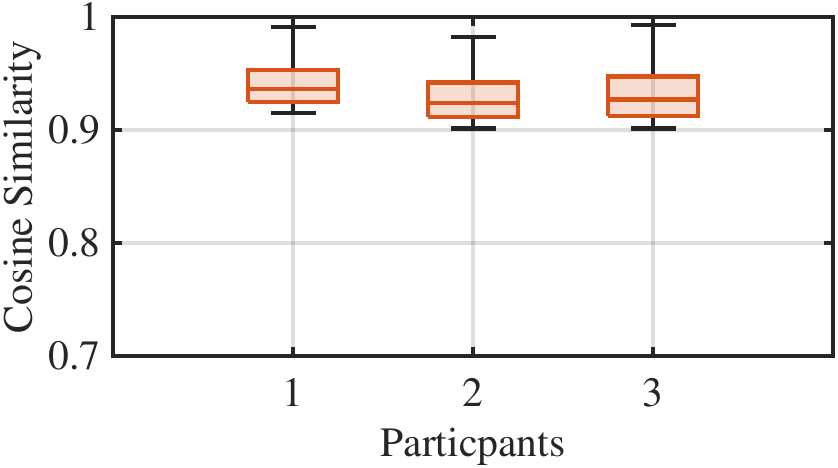}
       \vspace{-1ex}
       \label{fig:mutipleBPM}
     }
      \subfigure[Respiration Error]{
        \includegraphics[width=0.45\columnwidth]{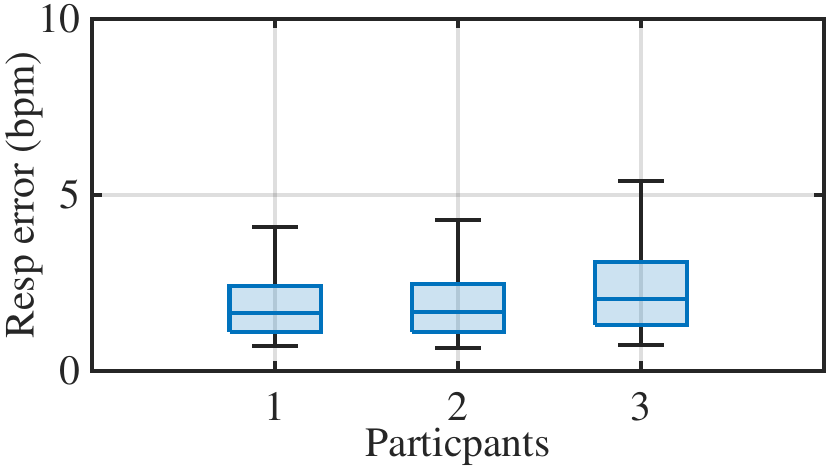}
        \vspace{-1ex}
        \label{fig:mutipleBPMNLOS}
      }
    \vspace{-1.5ex}
    \caption{Cosine similarity and BPM of respiration rate detection for multiple targets.}
    \label{fig:mutipleBPMComparison}
  \vspace{-2ex}
\end{figure}

\vspace{-0.8ex}
\subsubsection{Case of Multiple Targets}
\vspace{-0.5ex}
With multiple targets inside a room, \system can extract each individual's respiration waveforms by decomposing the composite RF signals. However, it cannot map these extracted waveforms to individual targets. Mapping the recovered individual waveforms to specific targets requires continuous localization and tracking of each target, which we leave as future work. Nonetheless, we show \system's ability to recover the individual respiration waveforms and to accurately measure the individual BPM by manually associating the recovered waveform to the target. Fig~\ref{fig:mutipleBPMComparison} shows the cosine similarity of the recovered waveforms for three mobile targets in a room and their BPM accuracy. Median cosine similarity for the three waveforms are 0.923, 0.934, and 0.9216, respectively, and median errors are 2.14 bpm , 2.23 bpm, and 2.57 bpm. 



\begin{figure}[htb]
\vspace{-0.5ex}
  \centering
  \includegraphics[width=0.9\columnwidth]{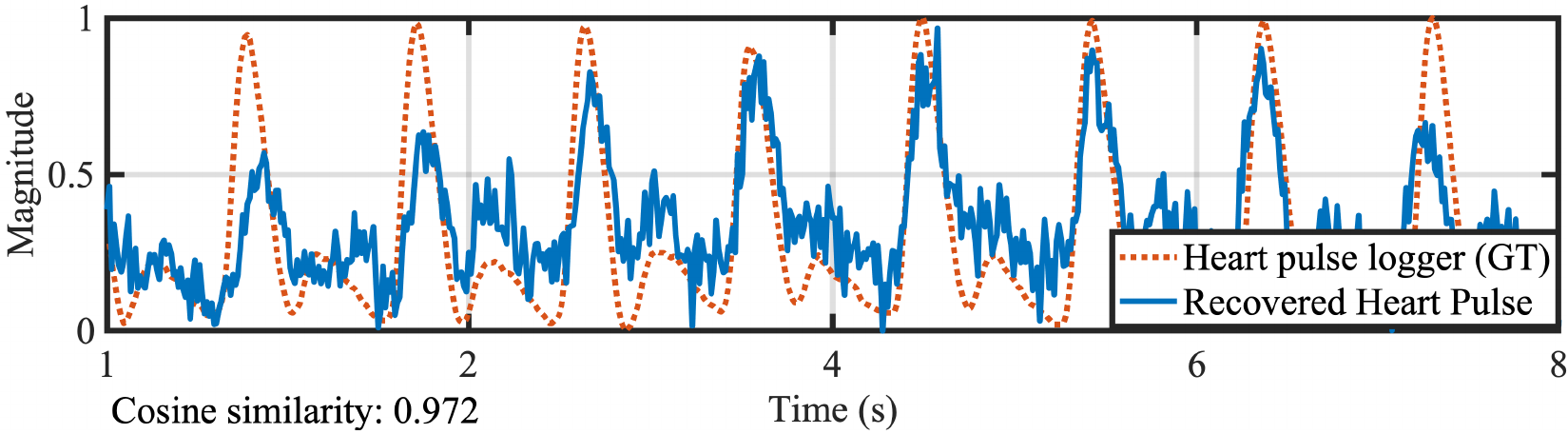}
  \vspace{-1ex}
  \caption{Recovered waveforms of heart pulses. Red dotted lines represent the ground-truth. Blue lines show the recovered waveforms of heart pulses by \system. }
  \label{fig:heartwaveform}
  \vspace{-2ex}
\end{figure}

\begin{figure}[htb]
    \vspace{-2.3ex}
     \centering
     \subfigure[Front]{
       \includegraphics[width=0.45\columnwidth]{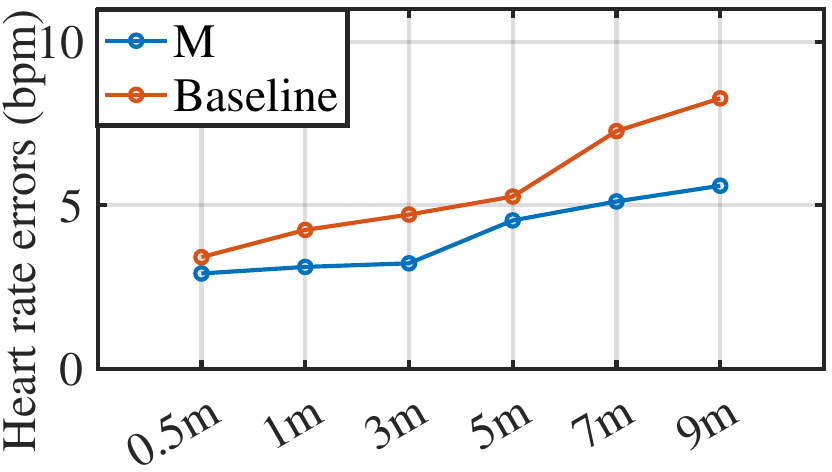}
       \vspace{-0.5ex}
       \label{fig:heartrate_front}
     }
     \subfigure[Left and Right]{
       \includegraphics[width=0.45\columnwidth]{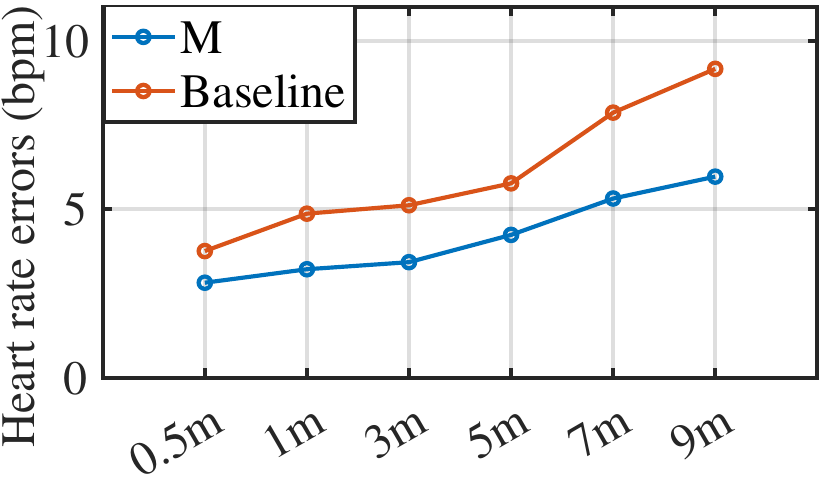}
       \vspace{-0.5ex}
       \label{fig:heartrate_front}
     }
     \vspace{-2.5ex}
     \caption{Heart rate error (bpm) measured over distances of 0.5–9m for static targets (SS) in various orientations of LoS}
    \label{fig:heartrate_static_los}
    \vspace{-2ex}
\end{figure}

\begin{figure}[htb]
    \vspace{-2.3ex}
     \centering
     \subfigure[Front]{
       \includegraphics[width=0.43\columnwidth]{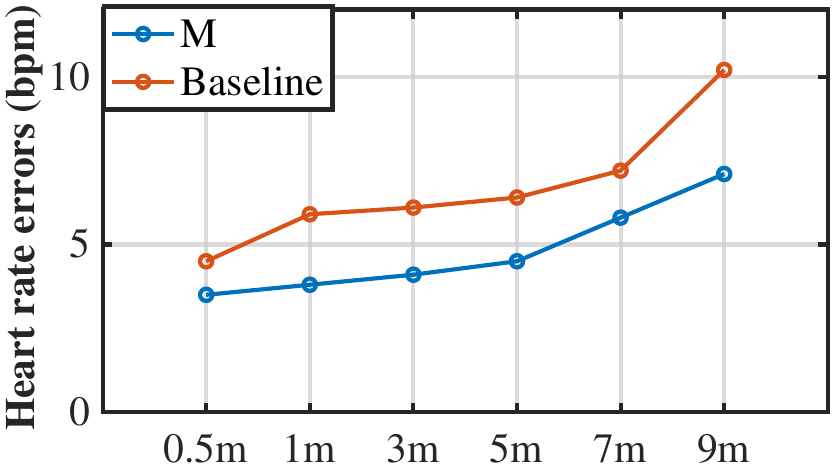}
       \vspace{-0.5ex}
       \label{fig:heartrate_front}
     }
     \subfigure[Left and Right]{
       \includegraphics[width=0.43\columnwidth]{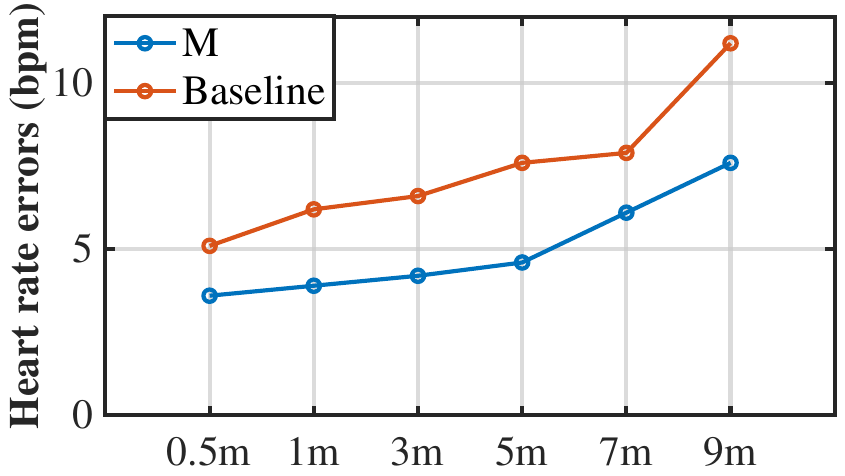}
       \vspace{-0.5ex}
       \label{fig:heartrate_front}
     }
     \vspace{-2.5ex}
     \caption{Heart rate error (bpm) measured over distances of 0.5–9m for static targets (SS) in various orientations of NLoS}
    \label{fig:heartrate_static_nlos}
    \vspace{-2ex}
\end{figure}

\begin{figure}[htb]
    \vspace{-2ex}
     \centering
     \subfigure[LoS]{
       \includegraphics[width=0.45\columnwidth]{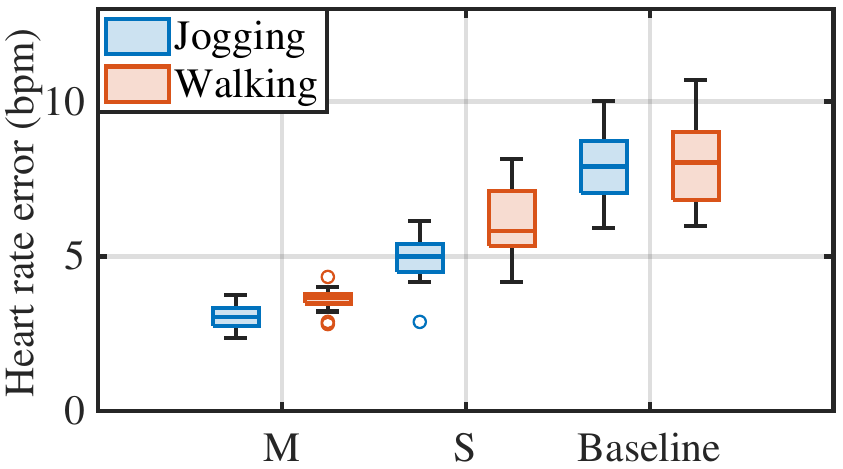}
       \label{fig:BPM_heartmove_los}
     }
     \subfigure[NLoS]{
        \includegraphics[width=0.45\columnwidth]{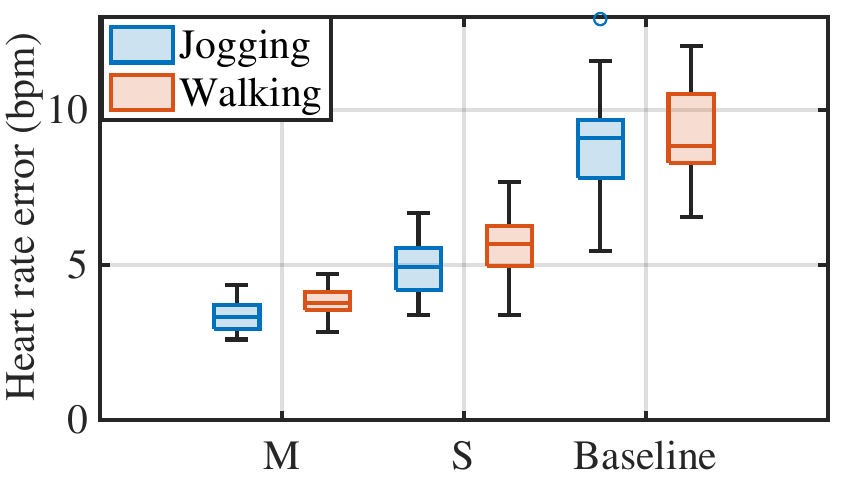}
        \label{fig:BPM_heartmove_nlos}
     }
     \vspace{-2ex}
     \caption{Targets with movement: Heart rate error (bpm) at 1 m distance in both LoS and NLoS conditions. Comparison of \system~(`M'), Co-located single-view MIMO radar~(`S'), and Baseline.} 
    \label{fig:BPM_heartmove}
    \vspace{-2.5ex}
\end{figure}

\vspace{-0.8ex}
\subsubsection{Measuring Heart Rate}
\vspace{-0.5ex}
Lastly, we demonstrate \system's heartbeat sensing capability.
%
To comprehensively evaluate heartbeat detection, participants were asked to perform random motions at varying distances. 
We compare our results against the COTS mmWave TI radar~\cite{TImmWave1443} used in previous work~\cite{ha2020contactless, gong_UbiComp2021} as our baseline. Figure~\ref{fig:heartwaveform} shows the recovered waveform from a static subject, while Figures~\ref{fig:heartrate_static_los} and \ref{fig:heartrate_static_nlos} present heart rate errors for static participants in various orientations under LoS and NLoS conditions, respectively. Although NLoS results (Fig.~\ref{fig:heartrate_static_nlos}) exhibit higher errors due to blockages, \system’s errors remain nearly constant, highlighting its robustness in complex environments.

At the closest 0.5m distance, \system offers the accurate estimation for both static subjects (3.22 bpm - front, 3.47 bpm - left and right) and subjects with motion (3.77 bpm), surpassing the baseline solution~(4.87 bpm). As distance increases, the performance of the baseline, which uses a single-view TI mmWave radar, significantly worsens compared to \system, with a widening discrepancy. The baseline achieved a heart rate error exceeding 10 bpm when tracking moving subjects at 9m, underscoring the limitations of single-view systems in practical scenarios.
For heart rate detection in jogging and walking experiments~\ref{fig:BPM_heartmove}, \system outperforms both the co-located single antenna array setup and the baseline, demonstrating clear advantages when subjects are in motion. Both the distributed (\system-‘M’) and co-located (\system-‘S’) configurations exceed the baseline in NLoS, highlighting the excellent penetration performance of UWB signals.

\vspace{-2ex}
\section{Discussion}
\vspace{-0.6ex}
\label{sec:discuss}
\textbf{\emph{Clinical Interpretation:}}While our evaluations focused mainly on healthy subjects, it is essential to contextualize our findings using clinical interpretation standards to determine if \system meets clinical requirements. 
%
Based on clinical studies~\cite{respiratory_imprecises}, 30-second respiratory rate measurements typically show an interquartile range of about 3.4 bpm, which decreases to roughly 2.5–3.0 bpm with longer sampling durations. An error range of 2.5–3.5 bpm is acceptable in clinical practice. In Section~\ref{sec:eval}, \system achieves median errors of 1.98 bpm for static targets, 2.12 bpm and 2.46 bpm for jogging and walking in LoS, and 2.32 bpm and 2.43 bpm for mobile targets in NLoS. These results indicate that the respiration rate errors produced by \system fall within clinically acceptable limits.
Clinical studies on heart rate measurement~\cite{heartrateaccuracy_2022}, including those using wrist-worn sensors~\cite{sensorbasedheartrateaccuracy_2017} during activities such as walking, running, and cycling, typically maintain errors within a 5\% threshold. Our results show that \system achieves error rates between 4.31\% and 4.65\%, which is near the acceptable limit. However, as distance increases, both baseline and \system errors exceed 5\% beyond 5m, falling outside clinical diagnostic standards. Considering that the 5\% threshold applies to wearable sensors that measure with no distance, the \system shows significant potential for robust heart rate detection.

\noindent \textbf{\emph{Deployment Cost:}} While our system achieves robust vital sign monitoring through distributed MIMO radar arrays, its deployment incurs significant costs due to the need for multiple synchronized hardware components and extensive calibration. In future work, we plan to reduce these costs by integrating and miniaturizing components, developing scalable connection architectures, and optimizing synchronization protocols to facilitate large-scale, real-world applications.

\vspace{-2ex}
\section{Conclusion}
\label{conclud}
\vspace{-0.5ex}
In summary, \system presents a novel multi-view 256-element virtual MIMO radar system that facilitates robust and precise distributed radar sensing across diverse target-environment configurations. Our comprehensive experiments demonstrate that \system significantly outperforms existing solutions in terms of vital-sign measurement accuracy for both stationary and moving targets. \system lays the foundation for a feasible, contact-free vital monitoring solution that can be effectively implemented at scale in real-world settings.
\vspace{-1.5ex}
\section*{Acknowledgement}
We thank our shepherd, Xia Zhou, and the anonymous reviewers for their feedback.
This work was supported in part by US National Science Foundation~(NSF) grants --- CNS-2112562, CNS-2107060, CNS-2208761, CNS-2213688, CNS-2312716, and the US Department of Commerce award 70NANB21H043.

\bibliographystyle{abbrv}
\bibliography{main}
\end{document}